\documentclass{aa} 
\usepackage{txfonts}

\usepackage{graphicx}
\usepackage{natbib}

\usepackage{morefloats}
\bibpunct{(}{)}{;}{a}{}{,} 
\usepackage{txfonts}
\begin{document}
  \title{Modeling the chemical evolution of a collapsing prestellar core in two spatial dimensions}
  \titlerunning{Modeling the chemical evolution of a collapsing prestellar core}

   \author{R.~J. van Weeren
          \inst{1}
          \and
           C. Brinch \inst{1,2}
          \and M.~R. Hogerheijde\inst{1}
          }

   \institute{Leiden Observatory, Leiden University,
              P.O. Box 9513, NL-2300 RA Leiden, The Netherlands,\\
              \email{rvweeren@strw.leidenuniv.nl}
         \and
             Argelander-Institut f\"ur Astronomie, Universit\"at Bonn, Auf dem H\"ugel 71, Bonn, D-53121, Germany\\
             }

   \date{Received February 15, 2008; accepted February 9, 2009}

 
  \abstract
   {The physical conditions in a collapsing cloud can be traced by observations of molecular lines. To correctly interpret these observations the abundance distributions of the observed species need to be derived. The chemistry in a collapsing molecular cloud is not in a steady state as the density and temperature evolve. We therefore need to follow chemical reactions, both in the gas phase and on dust grains, as well as gas-grain interactions, to predict the abundance distributions. 
 }
   {Our aim is to model the abundances of molecules, in the gas phase and on grain mantles in the form of ice, from prestellar core collapse to disk formation. We want to investigate the need for grain surface reactions and compare our results with observed abundances, column densities, and ice-mantle compositions. }
   {We use a 2-dimensional hydrodynamical simulation as a physical model from which we take the density, temperature, and the flow of the gas. Trace particles, moving along with the gas, are used to follow the chemistry during prestellar core collapse and disk formation.   We calculated abundance profiles and column densities for various species. The evolution of these abundances and the composition of ices on grain mantles were compared to observations and we tested the influence of grain surface reactions on the abundances of species. We also investigated the initial abundances to be adopted in more detailed modeling of protoplanetary disks by following the chemical evolution of trace particles accreting onto the disk.
   }
{Fractional abundances of \rm{HCO$^{+}$},  \rm{N$_{2}$H$^{+}$}, \rm{H$_{2}$CO}, \rm{HC$_{3}$N}, and \rm{CH$_{3}$OH} from our model with grain surface reactions provide a good match to observations, while abundances of \rm{CO}, \rm{CS}, \rm{SO}, \rm{HCN}, and \rm{HNC} show better agreement without grain surface reactions. The observed mantle composition of dust grains is best reproduced when we include surface reactions. The initial chemical abundances to be used for detailed modeling of a protoplanetary disk are found to be different from those in dark interstellar clouds. Ices with a binding energy lower than about 1200\noindent\mbox{ }K sublimate before accreting onto the disk, while those with a higher binding energy do not.
}
 {}

   \maketitle

\section{Introduction}
Star and planet formation can be studied through emission lines of molecules in the gas phase and from absorption lines of ices on dust grains. In star-forming regions, the chemistry is not in steady state as the physical conditions change throughout the process \citep{2004ApJ...617..360L}. The chemical processes in star-forming regions are still far from understood and give rise to a large number of different molecular species \cite[e.g.,][]{2003ApJ...593L..51C}. In particular, our knowledge about gas-dust interactions and the chemical reactions on grains surfaces is still very limited. However, we need to understand these processes if we want to correctly asses the physical conditions in star-forming regions and follow the chemical evolution from cold molecular cores towards stars and planetary systems.
  
Previous studies of the chemical evolution during star formation have (mainly) focused on prestellar core formation and contraction \citep[e.g.,][]{2005ApJ...620..330A}. Freeze-out of neutral species is an important process in these dark clouds resulting in chemical differentiation. This has been observed in numerous prestellar cores and dark clouds, e.g., \cite{2002ApJ...569..815T,2005A&A...429..181P}. These authors found that \rm{CO} and sulfur-bearing species (e.g., \rm{CS, SO}) deplete at larger radii (lower densities) while \rm{NH$_{3}$} and \rm{N$_{2}$H$^{+}$} deplete at significantly higher number densities of $n \gtrsim 10^{6}$ cm$^{-3}$. 
 
 \cite{2003ApJ...585..355R} modeled chemical abundances in the envelopes surrounding newborn stars gradually heated by the embedded protostar and \citet{2004ApJ...617..360L} followed the chemical evolution during the collapse of a Bonnor-Ebert sphere. Generally, they find a phase of depletion followed by sublimation due to the rising temperature in the envelope of the protostar. Different species sublimate at different radii from the central protostar due to differences in the binding energies of species onto dust grain mantles. These studies used one-dimensional time-evolving models for the density and temperature distribution. By following fluid elements or massless trace particles moving along with the radially in-falling gas, abundances could be derived as function of distance from the protostar. No surface reactions were included in both of these simulations. \cite{2006A&A...457..927G} presented a model for hot cores which included surface reactions. However, the underlying physical model was relatively simple: consisting of a phase of free-fall isothermal collapse and a phase of gradual heating of the gas and dust by a central Young Stellar Object (YSO). \cite{2008ApJ...674..984A} used the same chemical model as \cite{2006A&A...457..927G} but adopted the results of a one-dimensional radiation-hydrodynamics calculation as the physical model of the core. They found that organic species were mainly formed on grain surfaces at temperatures of $20-40$ K. These organic species then sublimated into the gas phase as the gas entered the inner hot ($T \gtrsim 100$ K) region surrounding the protostar.
   
 Separate studies have focused on the chemical evolution in static 2-dimensional protoplanetary disks \citep[e.g.,][]{1996ApJ...467..684A,1997ApJ...486L..51A,2001A&A...371.1107A}. In the cooler midplane of the disk various molecules freeze-out again and accumulate in ice mantles. Due to the strong radiation field from the protostar ion-molecule reactions become important in the upper disk layers and above \citep[e.g.,][]{1999osps.conf...97V,1998ASPC..132...54V,2004A&A...428..511J}.
  
 \cite{2002A&A...389..908J} derived \rm{CO} abundances, using Monte Carlo line radiative transfer modeling, for a sample of 18 low-mass protostars and prestellar cores.  An increase in the \rm{CO} abundance from Class 0 to Class I objects was observed \citep[classification according to ][]{1987IAUS..115....1L, 1987ApJ...312..788A,1993ApJ...406..122A}.  Class I objects showed almost normal cosmic fractional \rm{CO} abundances ($X_{\rm{CO}} =  n(\rm{CO})/n(\rm{H_{2}})$     $\sim2 \times 10^{-4}$, with $n(\mathrm{CO})$ and $n(\mathrm{H}_2)$ the number densities of \rm{CO} and \rm{H$_{2}$} respectively), while for Class 0 objects the abundances were almost an order of magnitude lower. \cite{2004A&A...416..603J,2005A&A...437..501J} expanded on this work and constrained the abundances for a range of species in the envelopes of the same 18 objects. They found various trends for the abundances of species as function of envelope mass, reflecting the fact that the chemistry is not in steady state during prestellar core collapse.  In general, it is thought that the envelopes of Class I objects have a lower density and higher temperature than Class 0 objects. When the temperature rises in the region surrounding the protostar this leads to the return of species, like \rm{CO}, by sublimation to the gas phase. At the same time, further out in the envelope the abundances also increase as the freeze-out timescale gets longer due to a lower density.
The picture emerging from these and other observations is that freeze-out and desorption are important processes for the evolution of chemical abundances from Class 0 to Class I objects. 
 
Molecular lines of depleted species in Class 0 and Class I objects can be accounted for with ``drop''-abundance profiles, \cite{2005A&A...435..177J}. In the inner and outer regions of the envelope there is no depletion (due to a high temperature, low density respectively), whereas at intermediate distances there is a drop in the abundance due to freeze-out. \rm{HCO$^{+}$} shows a clear relationship with \rm{CO}, while \rm{N$_{2}$H$^{+}$} abundances remain almost constant in the envelopes. Both of these effects can be explained by the formation and destruction channels of these species which both involve \rm{CO}. \rm{CH$_{3}$OH} abundances for the same sample of 18 objects from \citeauthor{2002A&A...389..908J} require in some cases abundance jumps in the inner regions.  \rm{H$_{2}$CO} abundances were best fitted with ``drop''-abundance profiles, although an inner jump for \rm{H$_{2}$CO} could not be ruled out for most sources. Abundances of complex species like \rm{CH$_{3}$OH} are not directly affected by the longer freeze-out timescales in the envelope because of their high binding energies. These species can only return (in large amounts) to the gas phase by grain destruction mechanism (e.g., in shocks) or by sublimation if the temperature is high enough ($T \gtrsim 60$ K, depending on the species).

The composition of ices on grain mantles is also of interest. Grain mantles provide surfaces where species can meet and react \citep[e.g.,][]{1977ApJ...212..396A}. Abundances of certain species, a typical example is \rm{CH$_{3}$OH}, can increase dramatically when compared to their formation in the gas phase \citep[e.g.,][]{2007A&A...467.1103G}.
Also, the mantle composition has an effect on the binding energy of species onto grains and thus couples to the gas-grain interactions and gas phase chemistry. Modeled ice fractions can be compared to the observed ice mantle compositions in order to constrain which surface reactions take place.

The evolution of chemical abundances during prestellar core collapse may also provide us with a so called ``chemical clock''. It is still unclear wether such a clock exists. For example the use of sulfur-bearing species as a chemical clock was investigated by \cite{2003A&A...399..567B}, but no definitive conclusion has been reached. Modeled abundances  can in principle  also be used to constrain hydrodynamical models of star formation if the chemistry is well enough understood. For example, high gas phase abundances of complex species formed on grain mantles, require the temperature to be at least higher than the sublimation temperatures of these species.

In \cite*{2008A&A...489..617B}, from now on called Paper {\rm I},  we followed for the first time the 2-dimensional time evolution of the \rm{CO} abundance during prestellar core collapse and disk formation.  A 2-dimensional hydrodynamical simulation was used to track the temperature, density, and follow particle motions. We also modeled line profiles for \rm{CO} and its isotopologue \rm{C$^{18}$O} using the line excitation and radiative transfer code RATRAN \citep{2000A&A...362..697H}. In particular, the anti-correlation of the fractional \rm{CO} abundance with envelope mass was reproduced. However, only a simple model for calculating the \rm{CO} abundance was used. In this second paper we present a method for modeling the full chemical evolution of a collapsing prestellar core in more than one dimension. We replace the model for \rm{CO} freeze-out/sublimation by a gas phase chemical network which also includes gas-grain interactions and grain surface reactions. This is necessary in order to derive the abundances of species which are more strongly influenced by gas phase or grain surface reactions. We apply this method to a collapsing one-solar mass rotating prestellar core with the physical conditions given by the 2-dimensional hydrodynamical simulation from Paper \rm{I}.  

Apart from the reasons mentioned above there are several important motivations for this work. First, with the upcoming higher resolution ($< 1\arcsec$) observations of ALMA it becomes important to extend the chemical modeling towards two and three dimensions as one-dimensional radial abundance distributions will no longer provide a good description of the observations. The resulting abundance distributions from our model can then be used in line excitation and radiative transfer codes to provide spatially resolved line profiles over the envelope and disk surrounding protostellar objects. This is needed to correctly derive the physical conditions around protostellar objects, such as the velocity field \citep*{2008A&A...489..607B}.

Second, our modeling also provides important constraints for the initial abundances of species to be adopted in more detailed disk modeling \citep[e.g.,][]{2004A&A...417...93S, 2006ApJ...647L..57S, 2007ApJ...669.1262P} as in our simulation we can track the chemical evolution of gas and ice accreting onto the disk (see Section \ref{sec:discussion}).

The method presented in this paper for modeling the chemical abundances is generally applicable and does not depend on the particular hydrodynamical simulation we have used. Our method can also handle one-dimensional as well as 3-dimensional hydrodynamical simulations as long as values for the density, temperature, velocity field, and the radiation field are provided.

The layout of this paper is as follows. In Section \ref{sec:allmodel} we shortly summarize the hydrodynamical simulation we use as input for our chemistry. We also present the chemical network to compute the gas phase abundances and ice composition. In Section \ref{sec:method} we describe our method to couple the output from the hydrodynamical simulation to the chemistry. In Section \ref{sec:results} we present our results and compare them with observations. This is followed by a discussion and conclusions in Section \ref{sec:discussion} and \ref{sec:conclusion}. 
\section{Model}
\label{sec:allmodel}
\subsection{Physical model: a 2-dimensional hydrodynamical simulation}
\label{hydro}
We use the same physical model of a collapsing prestellar core as in Paper I. The grid-based 2-dimensional simulation is described by \cite{1999ApJ...525..330Y}.  The simulation consists of multiple nested grids in order to sample areas with higher spatial resolution, e.g., the inner disk. The smallest scales correspond to $\sim7$ AU. The central cell is treated as a sink cell into which mass and angular momentum can flow \citep{1982ApJ...258..270B}. The Poisson equation for the gravitational potential is solved according to \cite{1975ApJ...199..619B}. For the hydrodynamics and radiation transport, the scheme from \cite{1985A&A...143...59R} is taken. Artificial viscosity is included for shocks. The scheme from \cite{1994ApJ...436..335L} is used for modeling angular momentum transport.

The initial conditions we adopt correspond to the J-model of \cite{1999ApJ...525..330Y}: a 1 M$_{\odot}$ isothermal sphere (a temperature of $10$ K for both the gas and dust) with an initial radius of $6667$ AU and a power law density profile with a slope of $-2$. The isothermal sphere was given a solid body rotational perturbation of $10^{-13}$ s$^{-1}$. For this set of initial conditions the free-fall time ($t_{\mathrm{ff}}$) is $1 \times 10^{5}$ yr. The simulation runs from $t = 0$  to $t =  2.5$ $t_{\mathrm{ff}}$.

\subsection{Chemical model}
\label{chem-model}

\subsubsection{Gas phase chemistry}
For computing the time-evolution of abundances of various species we make use of the UMIST 2007 database\footnote{http://www.udfa.net} of gas phase chemical reactions, \cite{2007A&A...466.1197W}. This updated network consist of 420 species linked by 4573 reactions. We use the RATE06  $dipole$ version of the network which has dipole-enhanced ion-neutral rate coefficients (applicable at low temperatures). The initial abundances are shown in Table \ref{tab:initial}.

\subsubsection{Gas-grain interactions}
We extend the gas phase chemical network with gas-grain interactions: freeze-out and desorption rates are calculated explicitly for each neutral species \citep[e.g.,][]{1995ApJ...441..222B}. Species frozen out create ice mantles on the dust grains. The depletion (freeze-out) rate, $k_{\mathrm{dep}}$, for neutral species can be written as
\begin{equation}
\label{freezeout1}
k_{\mathrm{dep}} = \pi a^{2} \bar{v} S n_{\mathrm{\mathrm{gr}}} \mbox{ ,}
\end{equation} 
with $a$ the mean grain radius, $\bar{v} = \sqrt{8k_{\mathrm{\mathrm{B}}}T_{\mathrm{gas}}/\pi m(X)}$ the mean thermal velocity of the gas, $S$ the sticking coefficient, $m(X)$ the mass of species $X$, and $n_{\mathrm{gr}}$ the grain number density. The sticking coefficient $S$ is a function of $T_{\mathrm{gas}}$ and $T_{\mathrm{dust}}$, depends on the species accreting, the velocity of the gas, the interaction energy between the grain surface and the gas species, and the excitation of the phonon spectrum of the grain. We ignore most of these complications and adopt a sticking probability of unity, except for atomic hydrogen for which  $S$ is described by a function of $T_{\mathrm{gas}}$ and $T_{\mathrm{dust}}$ \citep{2005pcim.book.....T}. Past studies have also considered lower values for the sticking coefficients \citep[e.g.,][]{1996ApJ...467..684A}. Lower limits for the sticking coefficients for \rm{CO}, \rm{O}$_{2}$, and \rm{N}$_{2}$ at $T_{\mathrm{dust}} = 15$\noindent\mbox{ } K were measured by \citet{2006A&A...449.1297B, 2007A&A...466.1005A}. The reported values were $0.9$, $0.85$, and $0.87$ respectively, with a typical error of $0.05$, so sticking coefficients which are significantly lower than unity do not seem to be justified, except at higher temperatures of $T \gtrsim 10^{2}$\noindent\mbox{ }K. Temperatures in our model are generally lower than this value so we choose not to lower the sticking coefficients. We take a ``standard'' mean grain radius of 0.1 $\mu$m, and a grain abundance of $1.33 \times 10^{-12}$ relative to $n_{\rm{H}} = n(\rm{H}) + 2n(\rm{H}_{2})$, with $n(\rm{H})$ the number density of atomic hydrogen, and $n(\rm{H}_{2})$ the number density of molecular hydrogen.

When a species ($X$) freezes out it becomes bound. The values for the binding energy, $E_{\mathrm{b}}(X)$, onto the surface are uncertain. Just a few have been measured in laboratories, examples are \rm{CO}, \rm{N$_{2}$}, \rm{O$_{2}$} \citep[see also][]{2004MNRAS.354.1133C}. The binding energies depend on the type of grain mantle (e.g,, bare \rm{SiO$_{2}$, \rm{CO} ice, \rm{H$_{2}$O} ice, etc.). To complicate things even further, grain mantles are mixtures of ices and evolve with time. Following \cite{2004ApJ...617..360L}, binding energies of a bare \rm{SiO$_{2}$} mantle are used as a starting point \citep[and references therein]{1992ApJS...82..167H,1993MNRAS.261...83H}. The binding energies onto an \rm{{H$_{2}$O} mantle (polar) and onto a \rm{CO} mantle (a-polar) are greater and smaller than that of a \rm{SiO$_{2}$} mantle by factors of 1.47 and 0.82, respectively. Note that these ratios are just rough estimates and could vary between species. Here we adopt \rm{CO} mantle binding energies following \cite {1996ApJ...467..684A}, i.e., we multiply the binding energies by a factor of $0.82$. For some species no binding energies were found in the literature. For species which are chemically similar to species with binding energies given, $E_{\mathrm{b}}$ is interpolated. Otherwise $E_{\mathrm{b}}$ is estimated as
 $E_{\mathrm{b}} = 50 A_{\mathrm{i}}$,
 with $A_{\mathrm{i}}$ the atomic weight \citep{2003A&A...399..197W} and $E_{\mathrm{b}}$ in units of Kelvin.

Species can be removed from grain mantles by various mechanisms. It is not completely clear yet which mechanisms operate and under what conditions, but at least thermal desorption (sublimation) is thought to be the most important one in protostellar environments.
The thermal desorption rate, $k_{\mathrm{sublm}}$, can be written as
\begin{equation}
\label{desorption}
k_{\mathrm{sublm}} = \nu(X) \exp\left(\frac{-E_{\mathrm{b}}(X)}{k_{\mathrm{B}} T_{\mathrm{dust}}}\right) \mbox{ ,}
\end{equation} 
with $\nu(X)$ the vibrational frequency of $X$ in its binding site, and $T_{\mathrm{dust}}$ the dust temperature. The vibrational frequency is the characteristic timescale for a species to acquire sufficient energy through thermal fluctuations in order to sublimate. With the harmonic oscillator approximation, the vibrational frequency of $X$ in its binding site is given by
 \begin{equation}
 \label{oscillator}
 \nu(X) = \sqrt{\frac{2 n_{\mathrm{s}} E_{\mathrm{b}}(X)}{\pi^{2} m(X)}}\mbox{ ,}
 \end{equation} 
with $n_{\mathrm{s}}$ is the surface density of adsorption sites on a grain. The value of $n_{\mathrm{s}}$ is estimated to be $\sim1.5 \times 10^{15}$ cm$^{-2}$ \citep{1992ApJS...82..167H}.

 A second desorption mechanism which is taken into account is cosmic ray induced desorption. This mechanism was first proposed by \cite{1972ApJ...174..321W}. Energetic nuclei might eject molecules from grain surfaces by either raising the temperature of the entire grain or by spot heating near the impact site. Cosmic ray heating is dominated by heavy ions because they deposit much more energy compared to lighter more abundant energetic nuclei. It is important in cold molecular cores as the temperatures there are too low for thermal desorption. The formulation of \cite{1993MNRAS.261...83H} is used to calculate the cosmic ray desorption rate
 \begin{equation}
 \label{desorption_cr}
 k_{\mathrm{cr}} = 3.16 \cdot 10^{-19} \nu(X) \exp\left(\frac{-E_{\mathrm{b}}(X)}{k_{\mathrm{B}} \cdot 70 \mbox{ [K]}}\right) \mbox{ .}
 \end{equation}

\subsubsection{Dissociative recombination on grain surfaces}
Under normal conditions inside a molecular cloud, most grains will be negatively charged \citep{1980PASJ...32..405U}. It is assumed that when a positive ion collides with a grain it will undergo dissociative recombination, with the same branching ratio as a dissociative recombination reaction in the gas phase. The products of these reactions return to the gas phase immediately. After the positive ion and the grain surface have been neutralized, an electron will stick immediately to the grain surface restoring its negative charge. If we assume that all grains have a negative charge then positive ions will collide more often with grains than their neutral counterparts. For single ionized species the collision rate is enhanced by a factor $C_{\mathrm{ion}}$. The rate coefficient for a dissociative recombination on a grain surface is then written as
\begin{equation}
\label{dissrec}
k_{\mathrm{dr}} = \sum_{i} \alpha_{i} \pi a^{2} \bar{v} n_{\mathrm{gr}} C_{\mathrm{ion}} \mbox{ ,}
\end{equation} 
with
\begin{equation}
\label{Cion}
C_{\mathrm{ion}} = S \left(1 + \frac{1.671 \times 10^{-3}}{a  T_{\mathrm{gas}}} \right) \mbox{ ,}
\end{equation} 
and $a$ and $T_{\mathrm{gas}}$ in cgs units \citep{1980PASJ...32..405U, 1992MNRAS.255..471R}. The sticking coefficient for all ions is taken unity. The summation in Eq. \ref{dissrec} is over all possible recombination channels for a single ion; the quantity $\alpha_{i}$ denotes the probability for a particular channel (branching ratio).

\subsubsection{Grain surface chemistry}
Chemical reactions between species on grain surfaces can be of considerable importance: for example the formation of \rm{H$_{2}$} on grain mantles is the dominant mechanism to form \rm{H$_{2}$}. For other possible grain surface reactions the situation is less clear but abundances of some species are hard to explain without surface reactions as the gas phase production is too low to account for the observed abundances \citep[see also][]{2005ESASP.577..205H}. We have extended the gas phase chemical network by grain surface reactions using the approach of \cite{1992ApJS...82..167H}.
 Two reactants wander over the surface of a dust grain until they find each other and (possibly) react. This wandering/diffusion can either happen by thermal hopping over energy barriers or by quantum mechanical tunneling (only efficient for light species).  The diffusion rate of species $i$ is given by the largest of the thermal hopping or tunneling rate
\begin{equation}
t_{i, \mathrm{thermal}} = \nu(X)\exp[E_{\mathrm{B}}(X)/k_{\mathrm{B}}T_{\mathrm{dust}}]
\end{equation}
\begin{equation}
t_{i,\mathrm{quantum}} = \nu(X)^{-1}\exp[(2a/\hbar)(2 m(X) E_{\mathrm{B}})^{1/2}] \mbox{ .}
 \end{equation}
For the energy barrier against diffusion, $E_{\mathrm{B}}$, we take a value of 30\% of the binding energy, i.e, $E_{\mathrm{B}} = 0.3E_{\mathrm{b}}$ \citep{1992ApJS...82..167H}.  The reaction rate between two species $i$ and $j$ is given by
 \begin{equation}
\label{srate}
k_{ij} = \kappa_{ij}(t_{i}^{-1} + t_{j}^{-1})/(N_{\mathrm{s}}n_{\mathrm{gr}}) \mbox{ ,}
\end{equation} 
with $N_{\mathrm{s}}$ the number of surface sites on a grain ($\sim10^{6}$), $\kappa_{ij} =\exp[-2(a_{\mathrm{t}}/\hbar)(2\mu E_{a})^{1/2}]$, with $a_{\mathrm{t}}$ the tunneling barrier (1 \AA), $\mu$ the reduced mass, and $E_{a}$ the activation energy for the reaction. For most reactions $E_{a} = 0$ so that $\kappa_{ij} = 1$. The set of surface reactions was taken from \cite{1992ApJS...82..167H,1993MNRAS.261...83H}, model N(4200K). This set of reactions was complimented by that of \cite{2002P&SS...50.1257C}, but including only the non-deuterated species. Activation energies for some specific reactions taken were:  $2500$ K for $\rm{H} + \rm{CO} \rightarrow \rm{HCO}$, $2500$ K for $\rm{H} + \rm{H_{2}CO} \rightarrow \rm{CH_{3}O}$, and $176$ K for $\rm{OH} + \rm{CO} \rightarrow \rm{CO_{2}} + \rm{H}$  \citep{2007A&A...469..973C}. For $ \rm{CO} + \rm{O} \rightarrow \rm{CO_{2}}$ we took $1000$ K \citep{1985A&A...152..130D}.

The surface rates were altered according to  \cite{2001A&A...375..673S} and \cite{2002P&SS...50.1257C} in order to cope with the discrete aspects of grain surface chemistry. This so called ``modified rate approach'' has been tested against a more detailed stochastic method by \cite{2004A&A...423..241S} and shows reasonable agreement with it.  In short, we compare the reaction rate coefficient (Eq. \ref{srate}) with that of the accretion and desorption rates. If the reaction rate is the smallest amongst the three the rate is not altered. In all other cases, the diffusion rate is replaced by the larger of the accretion and desorption rates. The net effect is to slow down the reactions which include at least one light species (i.e., atomic hydrogen). While not entirely correct, stochastic methods are still not available for large chemical networks and it is at least preferable the use the ``modified rate approach'' instead of the ordinary surface reaction rates.

\begin{table}
\begin{center}
\caption{The initial abundances}
\begin{tabular}{ll}
\hline
\hline
Species&Abundance$^{a}$\\
       &(Relative to the number of hydrogen nuclei)\\
\hline
H       & 0.0\\
H$_{2}$ & 0.5\\
He      & 9.75(-2)\\
O       & 1.80(-4)\\
C$^{+}$ & 7.86(-5)\\   
N       & 2.47(-5)\\
Mg$^{+}$& 1.09(-8)\\
Na$^{+}$& 2.25(-9)\\
Fe$^{+}$& 2.74(-9)\\
S$^{+}$ & 9.14(-8)\\
Cl$^{+}$& 1.00(-9)\\
P$^{+}$ & 2.16(-10)\\
F       & 2.00(-10)\\
Si$^{+}$& 9.74(-9)\\
e$^{-}$ & 7.87(-5)\\
Grains  & 1.33(-12)\\
\hline
\end{tabular}
\label{tab:initial}
\end{center}
$^{a}$ abundances are given as $a(b) = a \times 10^{b}$\vspace{2mm}

The initial abundances, most of them were taken from \cite{2001ApJ...552..639A}, the grain abundance was taken from \cite{1992ApJS...82..167H}. For the Fluorine abundance we take a fifth of Chlorine abundance \citep{1994ApJ...433..729Z}.
\end{table}

 \section{Method}
 \label{sec:method}
 \begin{figure}
\centering
\includegraphics[trim =0.3cm 0cm -0.3cm 0cm,width=0.5\textwidth]{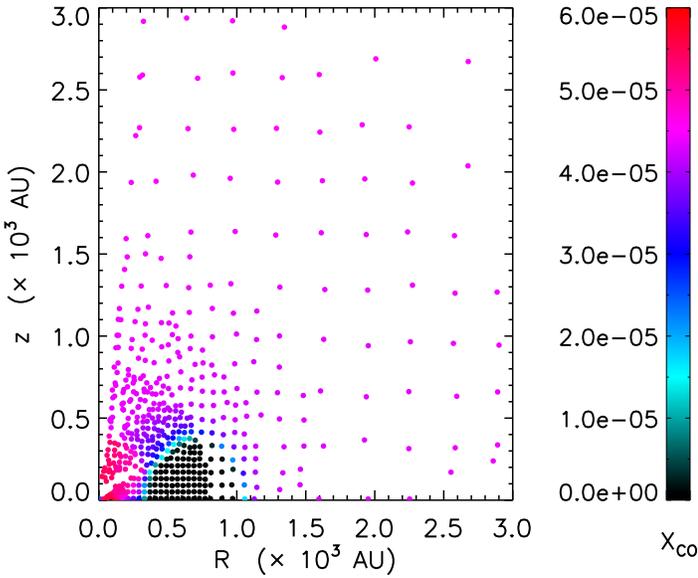} 
\caption{Distribution of trace particles at $t= 2.5$ $t_{\mathrm{ff}}$. The particle density increases inwards to sample these regions at a higher spatial resolution. In this example, the particles are color-coded according to the fractional \textrm{CO} abundance. For each of these trace particles we solve for the chemical abundances using Eq. \ref{eq:gas-phase} and  \ref{eq:surface}. The trace particles are more or less regularly distributed as the initial starting positions, $(R_{0},z_{0})$ at $t = 0$, were chosen such that at $t= 2.5$ $t_{\mathrm{ff}}$ they had moved to their respective location in order to properly sample the abundances throughout the core.}
\label{fig:dis}
\end{figure}

We integrate our initial abundances (Table \ref{tab:initial}) for $10^{5}$ yr using typical conditions for a dense interstellar medium (ISM): $n = 10^{4}$ cm$^{-3}$, $T_{\mathrm{gas}} = T_{\mathrm{dust}} = 10$ K, and an extinction corresponding to $A_{\mathrm{V}} = 5$. 
For this integration time, the gas phase abundances and grain mantle composition show a good match with observed abundances and ice-compositions in dark molecular clouds. The updated abundances then serve as starting abundances for the chemical simulation of the collapsing core.
Before the actual collapse we also take a static core configuration into account which lasts $3 \times 10^{5}$ yr (i.e., one dynamical timescale and roughly the lifetime of a prestellar core). The abundances are then further integrated corresponding to this static configuration, but now with the temperature and density given by the first snapshot ($t = 0$) of the hydrodynamical simulation. This approach is similar to that of \cite{2008ApJ...674..984A}. For direct cosmic-ray ionization reactions and cosmic-ray-induced photo-reactions we use the UMIST 2007 ionization rate $\zeta$ of $1.4 \times 10^{-17}$ s$^{-1}$ for \textrm{H$_2$}. For the photodissociation reactions we also use the rates as given by the UMIST 2007 database, corresponding to a standard unshielded interstellar ultraviolet radiation field of about $10^{8}$ photons cm$^{-2}$ s$^{-1}$ between 6 and 13.6 eV. 
For the extinction we take a constant value of $A_{\mathrm{V}} = 15$ throughout the core during the whole simulation which implies that the UV-radiation field has little effect on the chemistry. The cosmic ray ionization rate is left unaltered. 

The output of the hydrodynamical simulation serves as an input for the physical conditions needed in the chemistry. As in Paper {\rm I}, we populate the computational domain of the first snapshot ($t = 0$) with trace particles, see Fig. \ref{fig:dis}. The particles have no interaction with each other or the medium. The dynamics of these trace particles is discussed in Paper {\rm I}. Instead of solving for the \rm{CO} abundance only (using a very simple scheme), we compute abundances using a full set of chemical reactions (including gas-grain interactions and surface reactions). 

  \begin{figure}
\centering
\includegraphics[trim =0.3cm 0cm -0.3cm 0cm,width=0.5\textwidth]{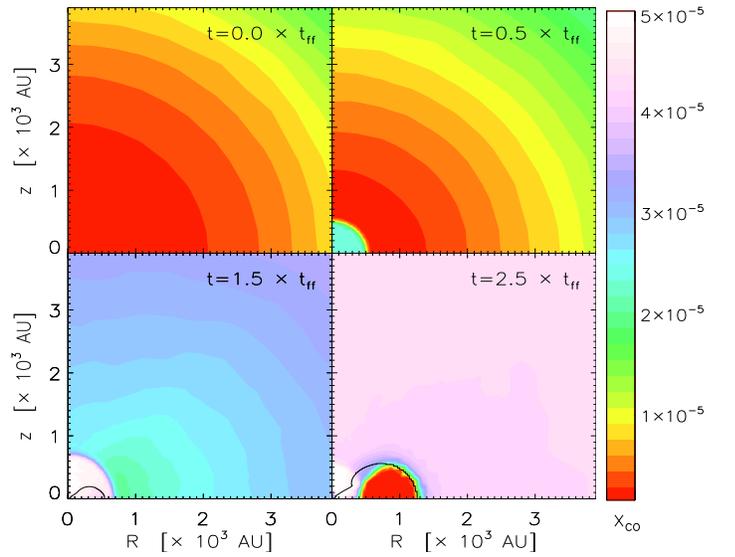} 
\caption{Distribution of CO throughout the core at four different times during the simulation for Model \rm{S}. The black contour shows the outline of the disk. The disk is defined as the region where $\sqrt{v_z^2 + v_R^2} \leq v_\phi$, with $v_\phi$ the rotational velocity.} 
\label{CO4x4}
\end{figure}
For each trace particle we solve the following differential equations for the gas phase fractional abundance $x(i)$, and surface fractional abundance $x_{s}(i)$ for species $i$
\begin{eqnarray}\label{eq:gas-phase}
\frac{dx(i)}{dt} = \sum_{l}  \sum_{j} K_{lj}x(l)x(j)n_{\mathrm{H_2}}    -  x(i) \sum_{j} K_{ij} x(j) n_{\mathrm{H_2}}\\
-k_{\mathrm{dep}}(i)x(i) +  [k_{\mathrm{sublm}}(i) + k_{\mathrm{cr}}(i)]x_{s}(i) \mbox{ ,} \nonumber
\end{eqnarray}
\begin{eqnarray}\label{eq:surface}
\frac{dx_{s}(i)}{dt} = \sum_{l} \sum_{j} k_{lj} x_{s}(l) x_{s}(j)n_{\mathrm{H_2}} -  x_{s}(i) \sum_{j} k_{ij}x_{s}(j)n_{\mathrm{H_2}}\\
+ k_{\mathrm{dep}}(i)x(i) - [k_{\mathrm{sublm}}(i) + k_{\mathrm{cr}}(i)]x_{s}(i) \mbox{ ,} \nonumber
\end{eqnarray}
with $K_{lj}$ the gas phase reaction rates between species $l$ and $j$ and $k_{ij}$ the grain surface reaction rate between species $i$ and $j$. In both equations the first and second terms correspond to the gas phase formation and destruction of the species respectively. The last two terms correspond to the depletion and desorption processes. Since the particles move with the flow of the gas the absolute abundances evolve according to the molecular hydrogen density ${n_{\mathrm{H_2} } \propto n(R, z, t)}$. The set of differential equations is solved using DLSODE \citep{DLSODE}. Thus, for a single trace particle the calculation goes as follows: a trace particle starts at a position ${(R_{0},z_{0})}$ within the core with the abundances found from our initial calculations. Rate coefficients (${K_{lj}}$ and ${k_{ij}}$) are calculated using the density and temperature at that position. The abundances in this trace particle (fluid element) are updated by integrating the differential equations of our chemical network (Eq. \ref{eq:gas-phase} and \ref{eq:surface}). The integration time is equal to the time between two neighboring snapshots of the hydrodynamical simulation. The position of the trace particle is then updated, with the velocities given by our 2-dimensional hydrodynamical model. This process is repeated until the end of the simulation ($\sim2100$ snapshots) or when the particle falls onto the protostar. We then proceed with the next trace particle.

We limit the number of particles to about 700 in order to carry out the computations in a reasonable amount of time. The trajectories of these 700 trace particles are a subset of that followed in Paper {\rm I}. Out of $9 \times 10^{5}$ particles, $700$ are chosen such that at particular time steps ($t = 0.0$, $0.5$, $1.5$, and $2.5$  $t_{\mathrm{ff}}$) these particles are well distributed spatially throughout the core with respect to the varying physical conditions (i.e., the density of the particles increases radially inwards to properly sample the inner regions). We thus follow 4 different subsets of about 700 particles each. The particles move with the flow of the gas so they are irregularly distributed with respect to their locations $(R,z)$. Therefore a mesh of triangles is constructed with data points (corresponding to the positions of the particles) at the vertices of the triangles. The mesh of triangles defines a piecewise-planar interpolating function, and we can thus grid the abundances onto a regular 2-dimensional grid. The spatial resolution varies from $\sim20$ AU in the core center to $\sim500$ AU in the outer envelope. 

In order to asses the effects of grain surface reactions on the abundances of species we run the same simulation again but without the grain surface reactions, referred to as Model \rm{G}.  The model which includes grain surface reaction we will refer to as Model \rm{S}.

 \section{Results}
 \label{sec:results}
 \subsection{Abundance profiles \& column densities}
\begin{figure}
\centering
\includegraphics[trim =0.cm 0cm -0.cm 0cm, width=0.5\textwidth]{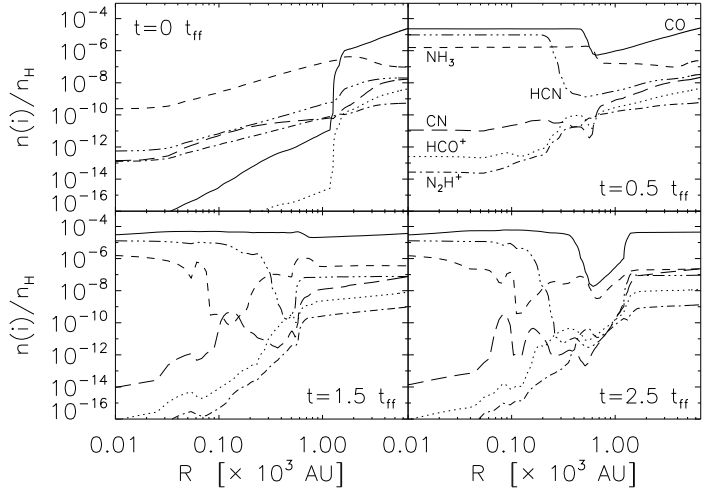} 
\caption{Model \rm{S}. The radial abundances through the disk ($R,z=0$) at four different times during the simulation. The line styles corresponding to different species are indicated in the top right panel at $t = 0.5$ $t_{\mathrm{ff}}$. Labeling is the same in the other three panels.} 
\label{radial_1}
\end{figure}
\begin{figure}
\centering
\includegraphics[trim = 0.0cm 0cm -0.0cm 0cm, width=0.5\textwidth]{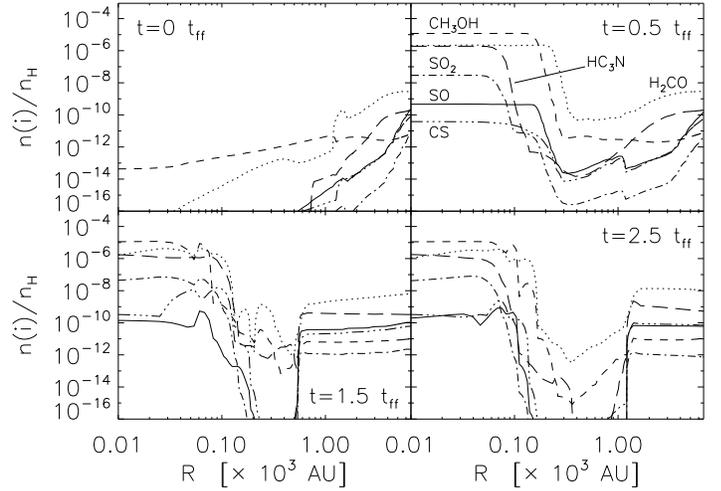} 
\caption{Model \rm{S}. The radial abundances through the disk ($R,z=0$) at four different times during the simulation. Labeling of species as in Fig. \ref{radial_1}. } 
\label{radial_2}
\end{figure}
\begin{figure}
\centering
\includegraphics[trim =0.cm 0cm -0.cm 0cm, width=0.5\textwidth]{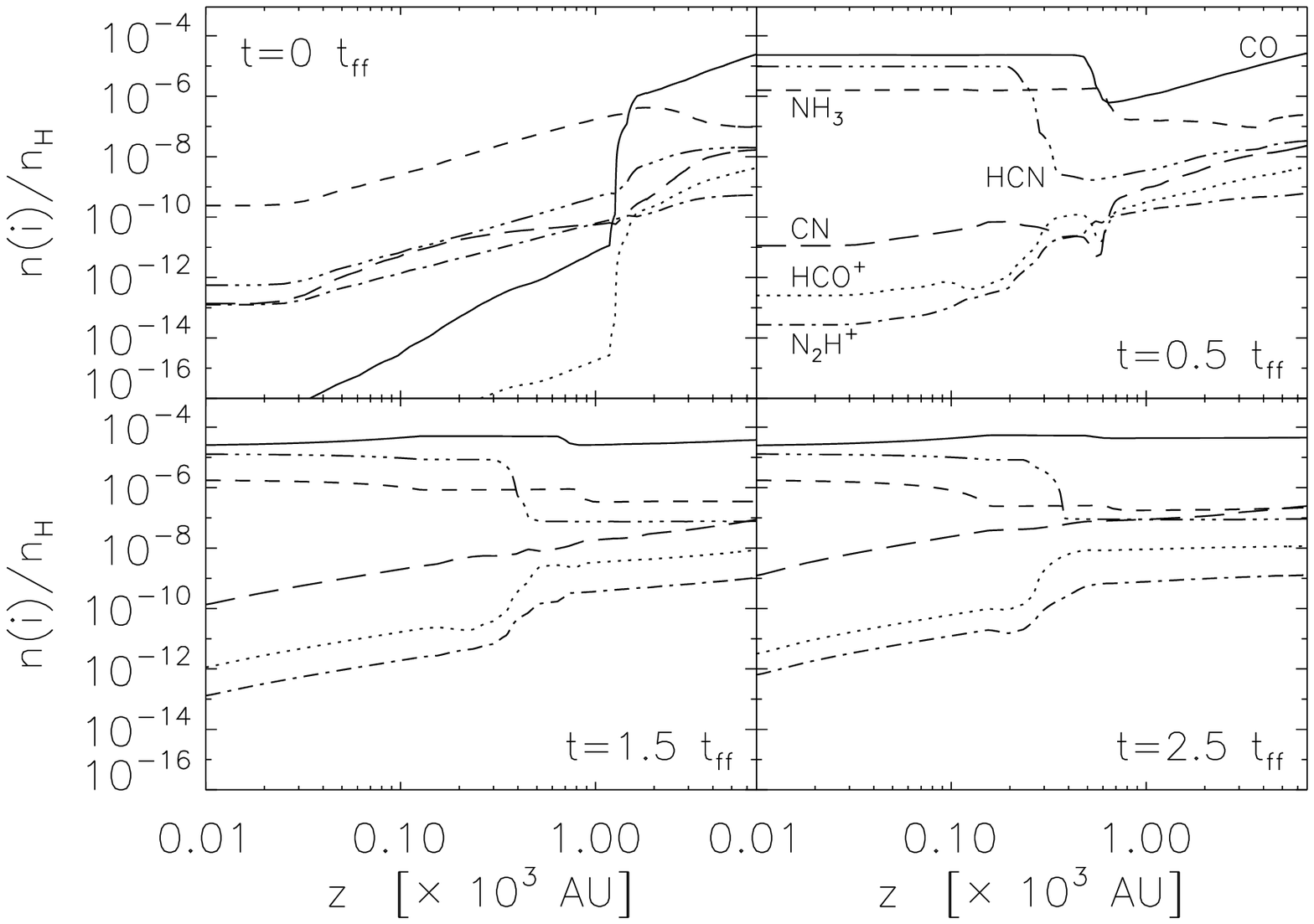} 
\caption{Model \rm{S}. The abundances perpendicular to the disk ($R=0,z$) at four different times during the simulation. Labeling of species as in Fig. \ref{radial_1}. } 
\label{polar_1}
\end{figure}
\begin{figure}
\centering
\includegraphics[trim =0.0cm 0cm -0.0cm 0cm, width=0.5\textwidth]{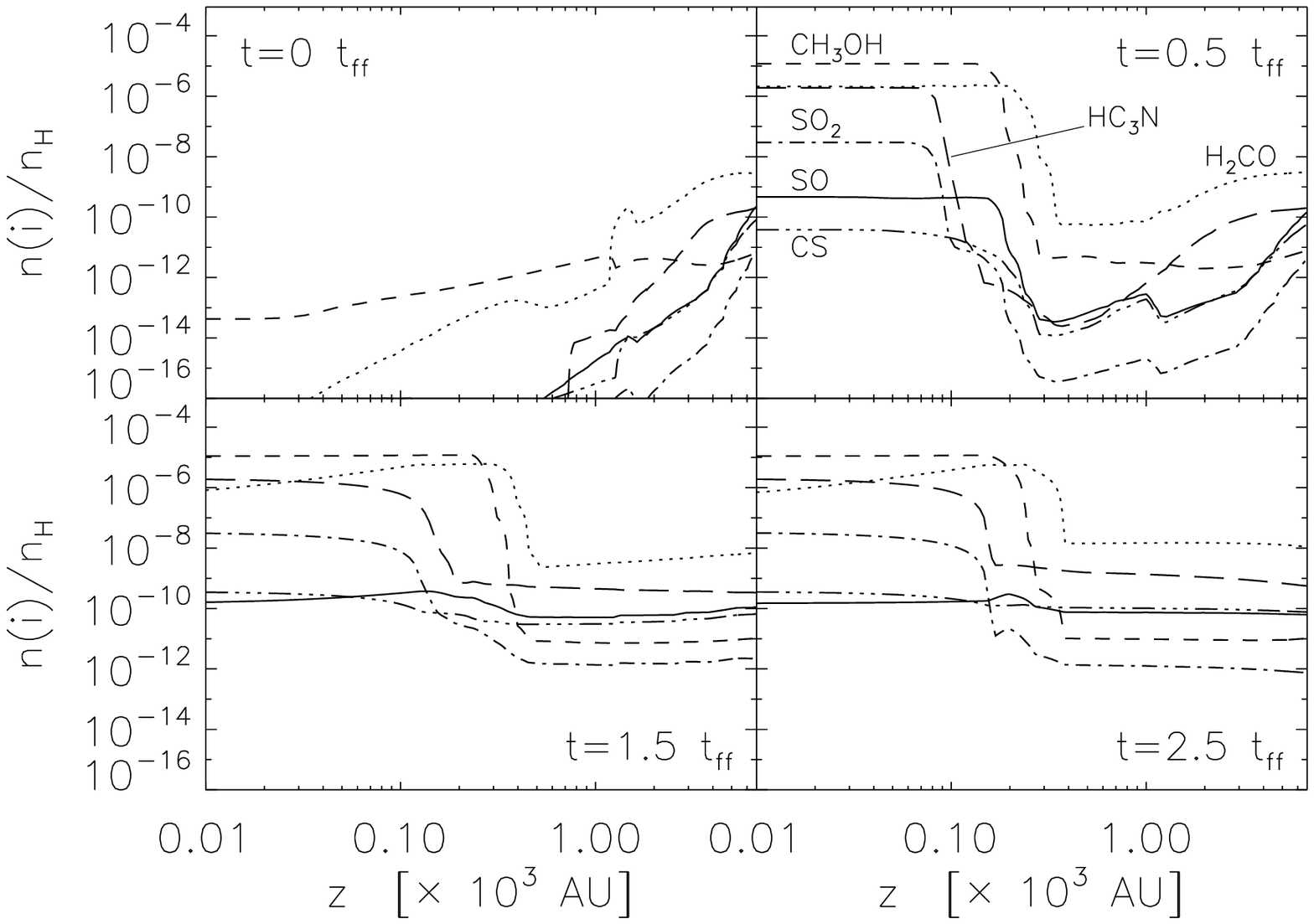} 
\caption{Model \rm{S}. The abundances perpendicular to the disk ($R=0,z$) at four different times during the simulation. Labeling of species as in Fig. \ref{radial_1}. } 
\label{polar_2}
\end{figure}
The 2-dimensional abundance distributions within the core for \rm{CO} (Model \rm{S}) are shown in Fig. \ref{CO4x4}.  We start with a centrally depleted core but as the temperature starts to rise in the center a sublimated zone develops. The depleted zone shrinks: as the cloud collapses the density in the envelope decreases so that the freeze-out rate drops. The sublimation radius is located at $\sim500$ AU. At the later stages \rm{CO} freezes out again in the disk due to the increase in density in this region. Qualitatively the results are similar to that in Paper \rm{I}, but the total \rm{CO} abundance of $5\times 10^{-5}$ is lower than the cosmic abundance of $2\times 10^{-4}$  used in Paper \rm{I} . This is due to the fact that \rm{CO} ice on grain mantles\noindent{ } efficiently transforms into other species (e.g, \rm{CO$_{2}$} and \rm{CH$_{3}$OH}) by grain surface reactions, so the amount of depletion seems higher. Model \rm{G} has a higher \rm{CO} abundance ($\sim10^{-4}$, similar to that in Paper I), because no surface conversion of \rm{CO} takes place, but apart from this scaling the abundance distribution looks like that in Fig. \ref{CO4x4}.

For other neutral species the pattern is similar, gas-grain interactions determine to a large extent the abundance evolution of species during prestellar core collapse. We start with a high level of depletion (several orders of magnitude locally), and when the temperature rises in the center a sublimated zone develops. In the outer parts of the envelope the density drops, resulting in the increase of the gas phase abundances of species with a relatively low binding energy. Within the disk, species with a low binding energy freeze-out again. Species with a higher binding energy ($\gtrsim 1200$ K) are able to enter the disk without sublimating. Gas accretes onto the disk around $\sim500$ AU (at the centrifugal radius) well outside the sublimation fronts for these species. Some ices (e.g., \rm{H$_2$O} and \rm{CH$_{3}$OH}) in the disk are thus (partly, as they are also being produced within the disk) of primordial origin. The abundances of charged species are determined by the balance between destruction and formation. Parent species, and/or destructors often have gas-grain interactions so the abundances of charged species are influenced by gas-grain interactions indirectly. The 2-dimensional abundance distributions for various species are included in the online Appendices \ref{appendix1} (Model \rm{S}) and \ref{appendix2} (Model \rm{G}).
\begin{figure}
\centering
\includegraphics[trim =0.3cm 0 -0.3cm 0cm,width=0.5\textwidth]{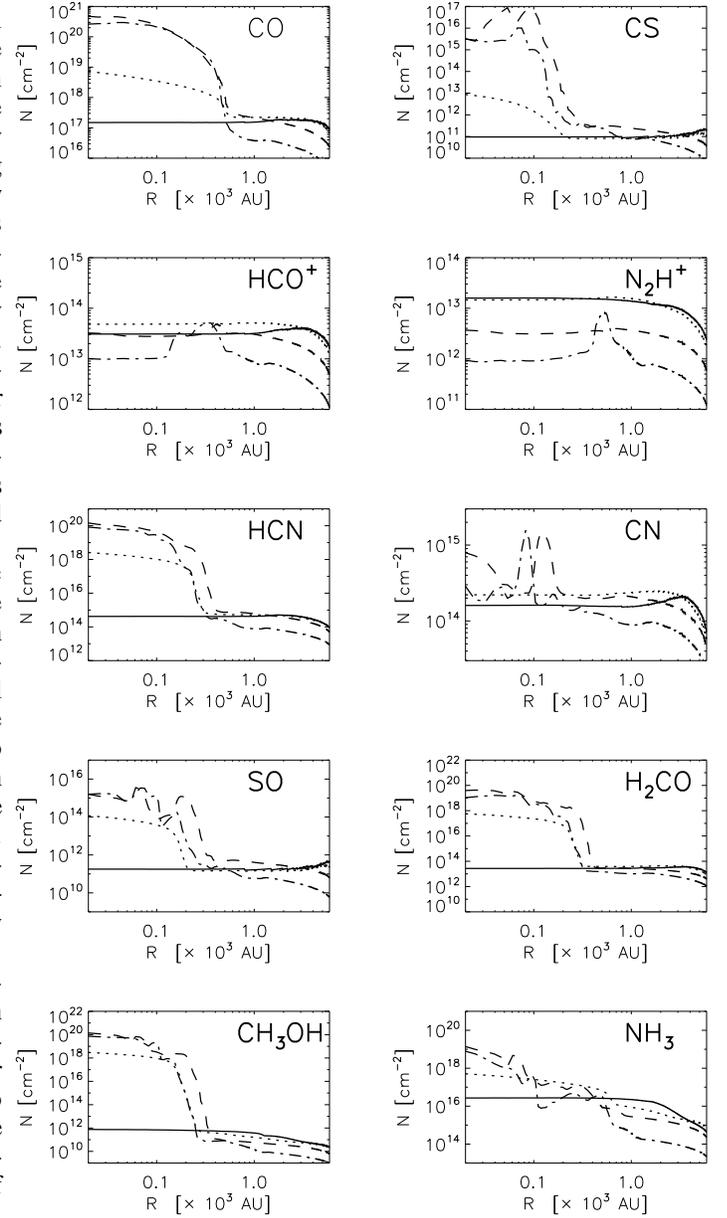} 
\caption{Column densities for various gas phase species at four different times during the simulation,  $t = 0$ (solid lines), $t = 0.5$ $t_{\mathrm{ff}}$ (dotted), $t = 1.5$ $t_{\mathrm{ff}}$ (dashed), and $t = 2.5$ $t_{\mathrm{ff}}$ (dash-dotted).} 
\label{N_gas}
\end{figure}

Abundances of various species as function of radius ($R,z=0$) are shown in Fig. \ref{radial_1} and \ref{radial_2}.  Abundances perpendicular to the disk plane ($R=0,z$) are shown in Fig. \ref{polar_1} and \ref{polar_2}. We have chosen to show only the results for Model \rm{S} as we will discuss the differences with Model \rm{G} in the next sections (with respect to global envelope abundances).
The radial profiles, cutting through the disk midplane, are difficult to compare with previous work due to the fact that the one-dimensional models do not include rotation and hence do not form a disk.  We can compare the abundance profiles perpendicular to the disk plane with that of \cite{2004ApJ...617..360L} and \cite{2008ApJ...674..984A}. 
The underlying physical models and chemical networks differ so that a one-to-one comparison of the results is not possible, instead we will try to identify differences or similar patterns in the abundance profiles.

We can see various abundance fluctuations for species especially at smaller radii ($\lesssim 1000$ AU) or near their sublimation front. In particular, species often show changes in the abundances near the \rm{CO} front, indicating that their formation or destruction is dependent on the \rm{CO} abundance, as was also noted by \cite{2004ApJ...617..360L}. For example the \rm{HCO$^{+}$} abundance increases near the \rm{CO} sublimation front, while the \rm{N$_{2}$H$^{+}$} abundance slightly decreases. Further inwards the abundances of these two species change again due to other gas phase reactions. Compared to \cite{2008ApJ...674..984A}, the \rm{NH$_{3}$} abundance shows a smaller increase at its sublimation front, about one order of magnitude compared to three orders of magnitude. This is caused by our lower binding energy of $887$ K compared to $5534$ K. We have chosen the lower value for the binding energy as we assume that the dust grains have an a-polar \rm{CO}-mantle.  This value may be correct at low temperatures (T $\lesssim 20$ K), but could be higher in the inner regions as most of the \rm{CO} mantle has sublimated and we are left with a mantle consisting mostly of \rm{H$_{2}$O} ice. The \rm{CS} abundances are two orders of magnitude lower then \cite{2008ApJ...674..984A}, while for \rm{HCN} they are higher by about the same amount. The \rm{HCO$^{+}$}, \rm{N$_{2}$H$^{+}$}, and \rm{H$_{2}$CO} abundances are similar. Note that when we compare the results for Model \rm{G} there is better agreement for some species, in particular for \rm{HCN} and \rm{CS}. The difference between Model \rm{S} and \rm{G} will be discussed in the next sections.

Compared to \cite{2004ApJ...617..360L}, \rm{CS}, \rm{HCN}, \rm{H$_{2}$CO}, and \rm{NH$_{3}$} abundance fluctuations are smaller, in particular at small radii and near sublimation fronts. As argued by  \cite{2008ApJ...674..984A}, these large fluctuations could arise from the fact that the number of species ($\sim80$) and chemical reactions ($\sim800$) are significantly lower compared to our network. In a small reaction network, the sudden change of the abundance of one species may easily change the abundance of other species. In larger networks, there are more species included in the formation and destruction channels of a particular species so a sudden change in the abundance of one species does not necessarily propagate towards other species.

We have calculated vertically integrated column densities for various species, ($N_{\rm{X}} = \int n_{\rm{X}}(z) \,dz$), which are shown in Fig. \ref{N_gas}. At $t = 0$, these species show flat profiles within $\sim1000$ AU. After the collapse neutral species show an increase in the column density at the sublimation front. For \rm{HCO$^{+}$}, there is an increase around $300$ AU (at $t = 1.5$ and $2.5$ $t_{\mathrm{ff}}$) but further inwards the column density decreases again and remains flat. For \rm{N$_{2}$H$^{+}$} the situation is similar. \rm{CN} shows relatively large variations in its column density inwards of $\sim200$ AU.
The column densities from \cite{2001ApJ...552..639A,2003ApJ...593..906A,2005ApJ...620..330A} can be compared to our first snapshot ($t = 0$). The results are comparable for most species, flat profiles towards the core center and slightly dropping for $R \gtrsim 5000$ AU. Detailed comparison is not possible as our underlying physical model is rather different. Column densities in the disk increase about three to four orders of magnitude as in \cite{1996ApJ...467..684A,2001A&A...371.1107A,2002A&A...386..622A} but are higher overall due to our massive disk of 0.4M$_{\odot}$.

 \subsection{Comparison with observations of envelope abundances}
 \cite{2002A&A...389..908J, 2004A&A...416..603J, 2005A&A...437..501J} presented a sample of abundances of various species in the envelopes of Class 0 and Class I objects, and in a few prestellar cores. Abundances were derived using observed line intensities and Monte Carlo radiative transfer modeling. In paper I, our \rm{CO} abundances show the same trend as \cite{2002A&A...389..908J}, i.e., a rising \rm{CO} abundance for a decreasing envelope mass. The global envelope abundance of a species in the simulation excludes the disk. The envelope is defined as the region where $\sqrt{v_z^2 + v_R^2} > v_\phi$, with $v_\phi$ the rotational velocity. The abundance is calculated by dividing the total number of molecules of a species in the envelope by the total number \rm{H$_{2}$} molecules in the envelope. We will now discuss the results for \rm{CO} and the other species which were present in the sample of \citeauthor{2002A&A...389..908J}. We have also included the results from Model G, which does not include grain surface reactions. By comparing the results of Model G and S we can determine the influence of surface reactions on the abundances. 
 
 \subsubsection{\rm{CO}}
The modeled \rm{CO} abundances as function of envelope mass are shown in Fig. \ref{XCOvsMenv}. The abundance increases from $7 \times 10^{-6}$ to $4 \times 10^{-5}$, for Model \rm{S}. Overall the \rm{CO} abundance is lower by about a factor of 10 compared to Model 3 in Paper I (including cosmic ray desorption and with $E_{\mathrm{b}} = 960$ K). Some of the Carbon (and Oxygen) ends up in other species than \rm{CO}. In particular, \rm{CO} on grain surfaces is transformed efficiently towards \rm{CO$_{2}$}, \rm{H$_{2}$CO}, \rm{CH$_{3}$OH} and other species. If the grain surface reactions are slowed down, we will get a better match to the observations. In fact, \rm{CO$_{2}$} ice and \rm{H$_{2}$CO} (gas) abundances are slightly overproduced in our simulation. The results of the simulation without grain surface reactions, Model \rm{G}, are very similar to that in Paper \rm{I}. This confirms that the \rm{CO} abundance is mostly determined by gas-grain interactions and grain surface reactions, and not by the gas phase chemistry during prestellar core collapse. 

\begin{figure}
\centering
\includegraphics[trim =0 0 0 0cm,width=0.45\textwidth]{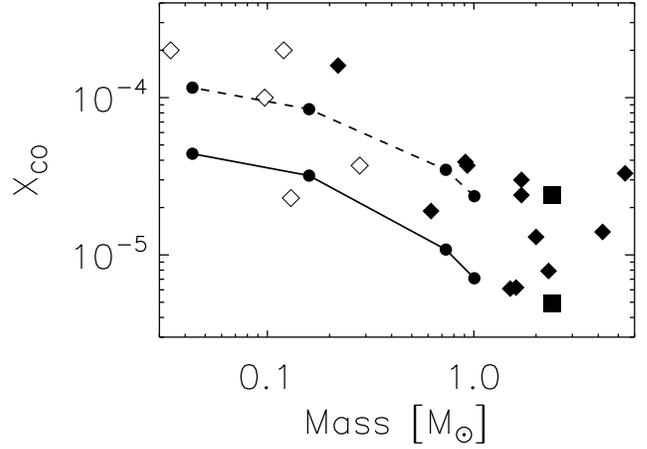} 
\caption{Global envelope abundances of \rm{CO} versus envelope mass. The bullets ``\textbullet'' connected by solid lines  are for Model \rm{S}, which includes grain surface reactions. The bullets connected by dashed lines are for Model \rm{G} without grain surface chemistry. Data for various objects from \cite{2002A&A...389..908J,2004A&A...416..603J} are overplotted. Class I objects are indicated by ``$\lozenge$'', Class 0 objects by ``$\blacklozenge$'', and prestellar cores by ``$\blacksquare$''.} 
\label{XCOvsMenv}
\end{figure}

\subsubsection{\rm{HCO$^{+}$}}

\begin{figure}
\centering
\includegraphics[trim =0 0 0 0cm,width=0.45\textwidth]{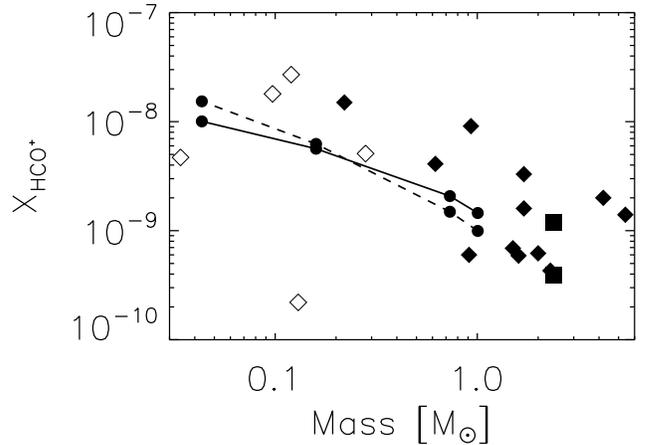} 
\caption{Global envelope abundances of \rm{HCO$^{+}$} versus envelope mass. Data for various objects from \cite{2004A&A...416..603J} are overplotted. Symbols and line styles are defined in Fig. \ref{XCOvsMenv}.
} 
\label{XHCO+vsMenv}
\end{figure}
 The main formation and destruction mechanisms for \rm{HCO$^{+}$} in the envelope are \citep{2004A&A...416..603J}
 \begin{equation}
 \label{HCOplusformation}
 \mbox{\rm{H$_{3}^{+}$}} +  \mbox{\rm{CO}} \rightarrow \mbox{\rm{HCO$^{+}$}} + \mbox{\rm{H$_{2}$}}
 \end{equation} 
 \begin{equation}
 \label{HCOplusdestruction}
 \mbox{\rm{HCO$^{+}$}} + \mbox{e$^{-}$} +  \rightarrow \mbox{\rm{CO}} + \mbox{\rm{H}} \mbox{ .}
 \end{equation} 
 Before collapse, the \rm{HCO$^{+}$} abundance decreases radially inwards. This is caused by freeze-out of its parent molecules (mainly \rm{CO}) and by dissociative recombination of \rm{HCO$^{+}$} (on grain surfaces). Within the sublimation front \rm{HCO$^{+}$} shows a decrease in its abundance at smaller radii because of a lower \rm{H$_{3}^{+}$} abundance. In Fig. \ref{XHCO+vsMenv} we have plotted $X_{\rm{HCO}^{+}}$ against envelope mass for 18 objects from \cite{2004A&A...416..603J}. Our evolution of the \rm{HCO$^{+}$} abundances agrees quite well with the observations, increasing from $10^{-9}$ to $10^{-8}$, indicating its chemistry is relatively well understood. The model without grain surface reactions shows a somewhat steeper evolution but overall is very similar, which also shows that the \rm{HCO$^{+}$} abundance is not greatly influenced by grain surface reactions.
\subsubsection{\rm{N$_{2}$H$^{+}$}}
The main formation mechanism for \rm{N$_{2}$H$^{+}$} \citep{2004A&A...416..603J} is:
\begin{equation}
\label{N2Hplusformation}
\mbox{\rm{H$_{3}^{+}$}} +  \mbox{\rm{N$_{2}$}} \rightarrow \mbox{\rm{N$_{2}$H$^{+}$}} + \mbox{\rm{H$_{2}$}} \mbox{ .}
\end{equation} 
The main destruction mechanisms are:
\begin{equation}
\label{N2Hplusdestruction1}
\mbox{\rm{N$_{2}$H$^{+}$}} + \mbox{e$^{-}$} \rightarrow \mbox{\rm{N$_{2}$}} + \mbox{\rm{H}}
\end{equation} 
\begin{equation}
\label{N2Hplusdestruction2}
\mbox{\rm{N$_{2}$H$^{+}$}} + \mbox{\rm{CO}} \rightarrow \mbox{\rm{N$_{2}$}} + \mbox{\rm{HCO$^{+}$}} \mbox{ .}
\end{equation} 
The \rm{N$_{2}$H$^{+}$} abundances decrease radially inwards before collapse. This is caused by freeze-out of its parent molecules (mainly N$_{2}$) and dissociative recombination of \rm{N$_{2}$H$^{+}$} (on grain surfaces). At $t=0$, the \rm{N$_{2}$H$^{+}$}} abundance drops slower inwards than the \rm{CO} abundance. Within the \rm{CO} sublimation front, \rm{N$_{2}$H$^{+}$} is destroyed by \rm{CO}, at later stages this  happens throughout the envelope. The low \rm{H$_{3}^{+}$} abundance in the center also slows down \rm{N$_{2}$H$^{+}$}} production. However, in the envelope the net abundance slightly rises due to a decreasing density.  Model \rm{G} shows lower abundances \citep[also noted by][]{2003ApJ...593..906A} and steeper evolution, both in disagreement with the observations. The higher abundances of about a factor two in the model with surface reactions are the result of the lower \rm{CO} abundance in combination with the effects of grain surface formation of \rm{N$_{2}$} and subsequent desorption of this parent molecule to the gas phase. We conclude that the model with grain surface reactions gives better results for \rm{N$_{2}$H$^{+}$}.

 \begin{figure}
 \centering
 \includegraphics[trim =0cm 0 0 0cm,width=0.45\textwidth]{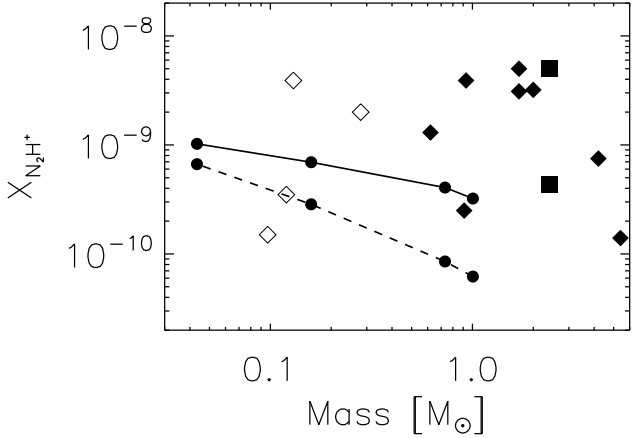} 
 \caption{Global envelope abundances of \rm{N$_{2}$H$^{+}$} versus envelope mass. Data for various objects from \cite{2004A&A...416..603J} are overplotted. Symbols and line styles are defined in Fig. \ref{XCOvsMenv}.
 } 
 \label{XN2H+vsMenv}
 \end{figure}

 \subsubsection{\rm{CS} and \rm{SO}}
The sulfur-bearing species \rm{CS} and \rm{SO} show relatively low abundances of $10^{-11} - 10^{-10}$ during the simulation. These molecules are heavily depleted before collapse due to relatively high binding energies of $1780$ K for both species. The abundances of both species increase by about an order of magnitude during the collapse due to the lower density (outer envelope) and higher temperature (center). For \rm{SO}, in the center a spherical central sublimated zone develops ($t = 0.5$ $t_{\mathrm{ff}}$) with an abundance of $\sim 4 \times 10^{-10}$. Above the midplane of the disk at a radius of $\sim200$ AU the abundance rises towards $\sim 3 \times 10^{-9}$. For \rm{CS} the pattern is similar, an abundance of $4 \times 10^{-11}$ in the center at \noindent{$t = 0.5$ $t_{\mathrm{ff}}$}, the abundance peaks around $t = 1.5$ $t_{\mathrm{ff}}$ at $1 \times 10^{-8}$, above and within the disk at $\rm 100$ AU. The enhancement is however too localized to have a major influence on the global envelope abundances. Observed global abundances are more then an order of magnitude higher and the abundances for Model \rm{G} agree with the observations. This difference could be caused by grain surfaces reactions converting both species towards other species. If we would lower the rates for these reactions surface abundances of \rm{CS} and \rm{SO} increase, and as a result gas phase abundances should increase as well. The increase or decrease in the abundances of other species, linked to reaction networks of \rm{CS} and \rm{SO}, could also have an effect. Abundances of both species can also be increased by the impact of outflows which processes the gas \citep[e.g.,][]{1997ApJ...487L..93B}.  As outflows are not included we could underestimate the abundances.

It has been suggested that sulfur-bearing species might provide a chemical clock \citep[e.g.,][]{1999MNRAS.306..691R} . In the case of low mass protostars this was further investigated by \cite{2003A&A...399..567B}. The chemical clock works as follows: large amounts of \rm{H$_{2}$S} are released from grain mantles by the rising temperature in the inner envelope.  \rm{H$_{2}$S}  is converted via several reactions steps into \rm{SO} and \rm{SO$_{2}$} on timescales of $10^{4} - 10^{5}$ years. The abundances of these species drop at later times as they are converted into \rm{CS}, \rm{H$_{2}$CS}, and \rm{OCS}. Model \rm{S} shows such a trend, at later times the \rm{SO} abundance slightly drops and the \rm{CS} abundance continues to increase. By comparing the abundances of species like \rm{CS}, \rm{SO},  \rm{SO$_{2}$}, \rm{H$_{2}$S}, and \rm{H$_{2}$CS} the timescale since the release of  \rm{H$_{2}$S} can hopefully be constrained. \cite{2003A&A...399..567B} found that to reproduce the observed abundances the initial \rm{H$_{2}$S} abundance should be $> 10^{-8}$. They did not compute the initial abundance of  \rm{H$_{2}$S}. Instead, they estimated the abundance from observed gas phase and ice abundances, since no grain surface chemistry was included in their model. In our simulation the surface abundance of \rm{H$_{2}$S} is $\sim 7 \times 10^{-8}$ before collapse in agreement with their estimates that X$_{\rm{H_{2}S}} > 10^{-8}$, although the abundance is on the low side. However, observed abundances of \rm{SO} \citep{2004A&A...416..603J} do not confirm the picture described above, i.e., they did not find systematically higher \rm{SO} abundances for Class 0 objects compared to Class I objects. We do also not see such an effect in our model without grain surface reactions. More observations of various sulfur-bearing species are needed to determine if they could be used as a chemical clock.

\begin{figure}
\centering
\includegraphics[trim =0 0 0 0cm,width=0.45\textwidth]{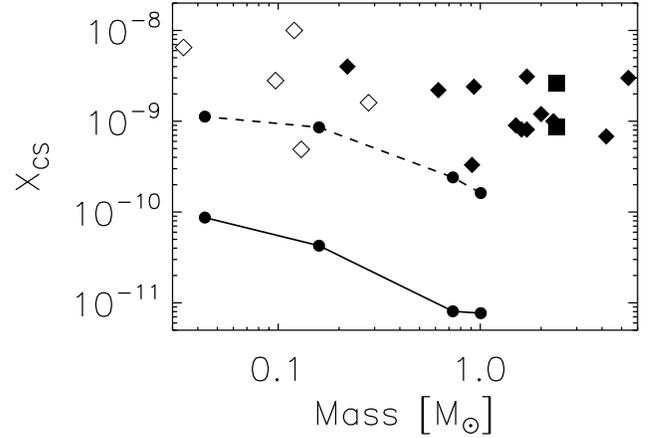} 
\caption{Global envelope abundances of CS versus envelope mass. Data for various objects from \cite{2004A&A...416..603J} are overplotted. Symbols and line styles are defined in Fig. \ref{XCOvsMenv}.
} 
\label{XCSvsMenv}
\end{figure}

\begin{figure}
\centering
\includegraphics[trim =0 0 0 0cm,width=0.45\textwidth]{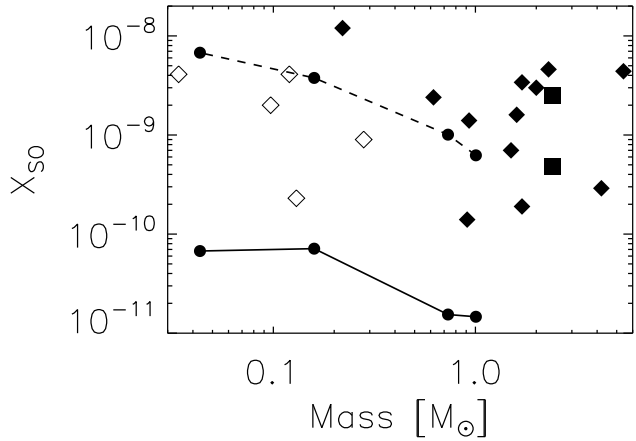} 
\caption{Global envelope abundances of \rm{SO} versus envelope mass. Data for various objects from \cite{2004A&A...416..603J} are overplotted. Symbols and line styles are defined in Fig. \ref{XCOvsMenv}.
} 
\label{XSOvsMenv}
\end{figure}

\subsubsection{\rm{HCN}, \rm{HNC} and \rm{CN}}
\label{sec:CN}
\rm{HCN} and \rm{HNC} abundances observed by \cite{2004A&A...416..603J} do not show a clear correlation with envelope mass. \rm{CN} abundances increase with decreasing envelope mass. This trend for \rm{CN} is also apparent in our simulation, although abundances are an order of magnitude too high compared to the observations. \rm{CN} abundances are too low by one order of magnitude for our model without grain surface reactions. \rm{HCN} and \rm{HNC} abundances  are very  similar and vary from $10^{-8} - 10^{-7}$. Initially, abundances rise, at $t \gtrsim 1.5$ $t_{\mathrm{ff}}$ the abundances remain more or less constant. \rm{HCN} abundances are an order of magnitude too high, \rm{HNC} abundances around two orders of magnitude. Both are produced in the gas phase but also by grain surface reactions. 

The gas phase production of \textrm{HCN},  \textrm{HNC}, and \textrm{CN} depends on the abundance of nitrogen-bearing species. Several critical reactions involving nitrogen-bearing species have large uncertainties in the rate coefficients, caused by the lack of theoretical or experimental verification of the existence of reaction barriers \citep[e.g.,][]{2006A&A...456..215F}.  Furthermore, the elemental C:O abundance ratio as well as the ${n(\mathrm{N})/n({N}_2)}$ ratio has an effect on the nitrogen-bearing chemistry as was shown by these authors. Another important aspect is the sticking coefficient of atomic nitrogen and oxygen. \cite{2006A&A...456..215F, 2007A&A...462..221A} found that sticking coefficients lower than unity for these species could explain the relatively high abundances of  \textrm{N$_{2}$H$^{+}$} and \textrm{NH$_3$} in the gas phase when other molecules, such as \textrm{CO}, have already been frozen out. We conclude that the surface rates involving the production of \textrm{HCN},  \textrm{HNC}, and \textrm{CN} should probably be decreased, but that the uncertainties in the gas phase nitrogen-bearing chemistry may also have caused our abundances to differ from the observed values. 

\begin{figure}
\centering
\includegraphics[trim =0 0 0 0cm,width=0.45\textwidth]{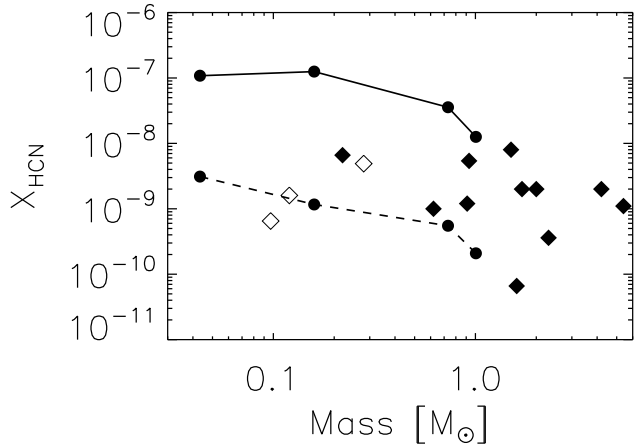} 
\caption{Global envelope abundances of \rm{HCN} versus envelope mass. Data for various objects from \cite{2004A&A...416..603J} are overplotted. Symbols and line styles are defined in Fig. \ref{XCOvsMenv}.
} 
\label{XHCNvsMenv}
\end{figure}

\begin{figure}
\centering
\includegraphics[trim =0 0 0 0cm,width=0.45\textwidth]{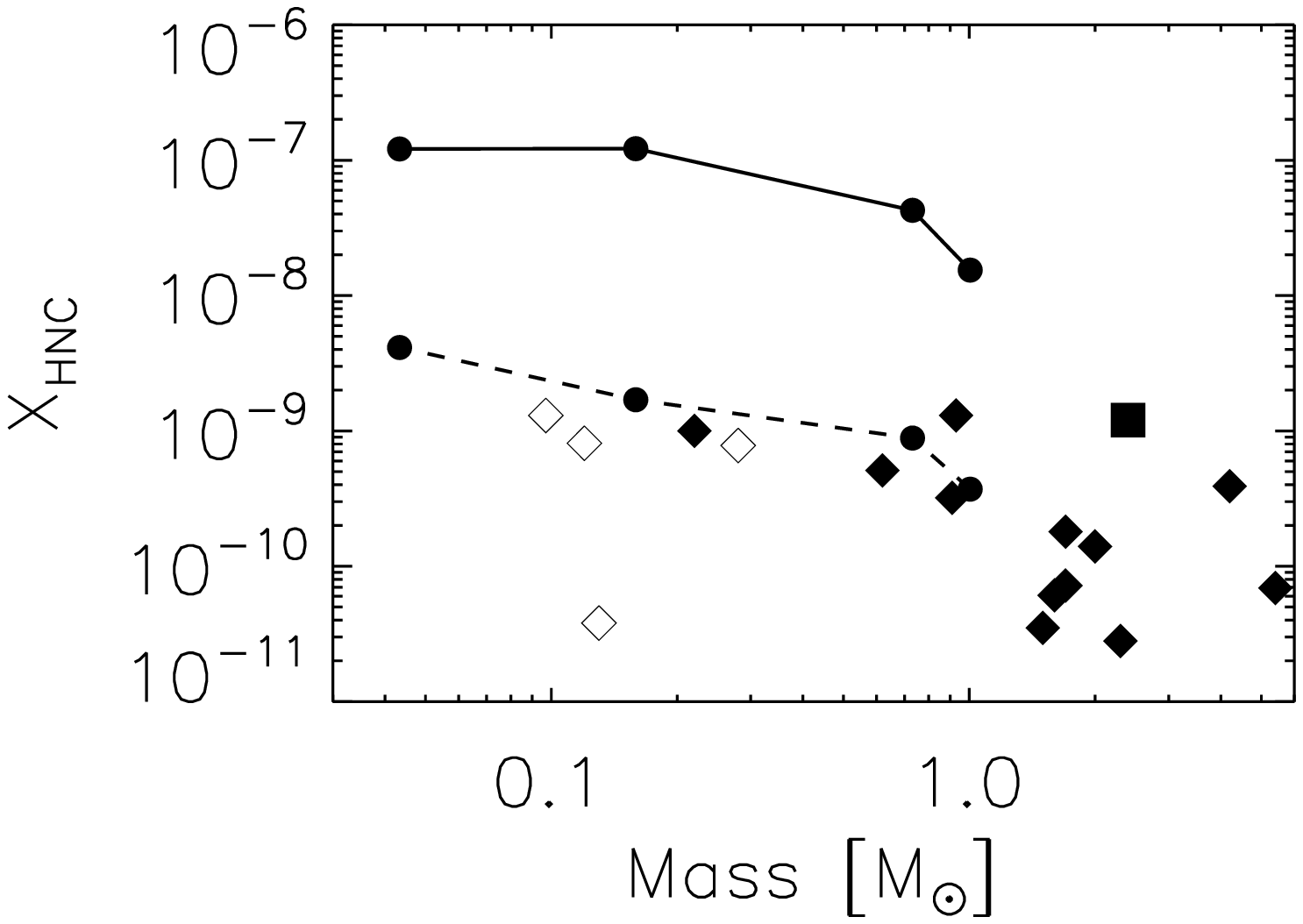} 
\caption{Global envelope abundances of \rm{HNC} versus envelope mass. Data for various objects from \cite{2004A&A...416..603J} are overplotted. Symbols and line styles are defined in Fig. \ref{XCOvsMenv}.
} 
\label{XHNCvsMenv}
\end{figure}

\begin{figure}
\centering
\includegraphics[trim =0 0 0 0cm,width=0.45\textwidth]{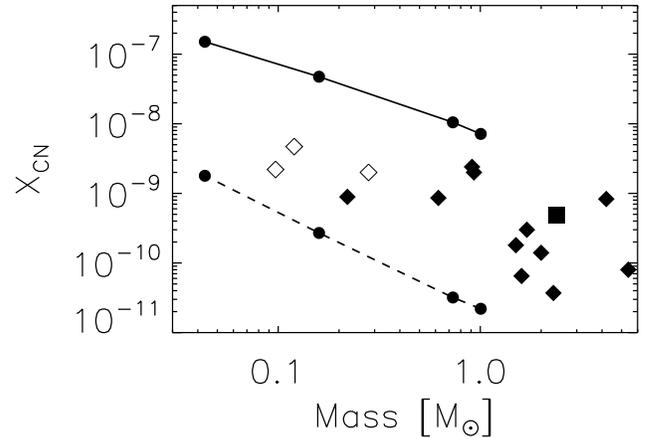} 
\caption{Global envelope abundances of \rm{CN} versus envelope mass. Data for various objects from \cite{2004A&A...416..603J} are overplotted. Symbols and line styles are defined in Fig. \ref{XCOvsMenv}.
} 
\label{XCNvsMenv}
\end{figure}

\subsubsection{\rm{HC$_{3}$N}}
Our modeled \rm{HC$_{3}$N} abundances provide a relatively good match to observed values, rising from $10^{-10}$ to $10^{-9}$. Without surface reactions, abundances drop by about two orders of magnitude. Some chemical models predict an increase in the abundance when the amount of depletion of other gas phase species (like \rm{CO}) is high \citep[e.g.,][]{1992ApJ...394..539H,1997MNRAS.291..235R,1998ApJ...499..234C}. Observed abundances do not show such a trend as does our model, in fact our model predicts slightly higher abundances with decreasing envelope mass due to sublimation of \rm{HC$_{3}$N} from grain mantles and lower freeze-out rates. \cite{2004A&A...416..603J} mention the possibility (amongst others) that the UV (ultraviolet) radiation from the central protostar could also lead to such a trend. Our results do not support this explanation as we have not included the UV radiation from the central star. In fact, \rm{HC$_{3}$N}  produced on grain surfaces and subsequently released to the gas phase could produce the observed values.
\begin{figure}
\centering
\includegraphics[trim =0 0 0 0cm,width=0.45\textwidth]{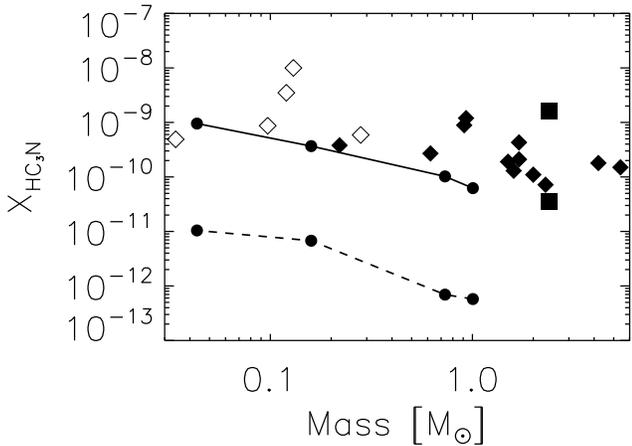} 
\caption{Global envelope abundances of \rm{HC$_{3}$N} versus envelope mass. Data for various objects from \cite{2004A&A...416..603J} are overplotted. Symbols and line styles are defined in Fig. \ref{XCOvsMenv}.
} 
\label{XHC3NvsMenv}
\end{figure}

\subsubsection{\rm{H$_{2}$CO} and \rm{CH$_{3}$OH}}
Observed \rm{H$_{2}$CO} abundances \citep{2005A&A...437..501J} show a clear anti-correlation with the envelope mass. \rm{H$_{2}$CO} freezes out at high densities and low temperatures. Sublimation takes places as the temperature rises in the inner regions and the freeze-out rate drops as the density in the envelope decreases. Our modeled abundances confirm this picture. The model without grain surface reactions produces abundances about an order of magnitude lower.

Our modeled \rm{CH$_{3}$OH} abundances show large variations, in the prestellar core observed gas phase abundances are very low ($10^{-12} - 10^{-11}$) because of the large binding energy. Due to the rising temperature and subsequent sublimation, the \rm{CH$_{3}$OH} abundance quickly increases in the center, at later times the abundance drops again. The model without surface reactions fails to reproduce the observed abundances by about three to four orders of magnitude. Most of the \rm{CH$_{3}$OH} is produced by grain surface reactions, where \rm{CO} is converted via several steps towards \rm{CH$_{3}$OH} which then sublimates into the gas phase. It is also believed that \rm{CH$_{3}$OH} abundances and to a lesser extend \rm{H$_{2}$CO} are influenced by other processes such as the impact of outflows \citep[e.g.,][]{1997ApJ...487L..93B}. \rm{CH$_{3}$OH} was only observed towards a handful of objects by \cite{2004A&A...416..603J}. This makes comparison with models difficult. At least modeled abundances are roughly within the range of observed values. Increasing the number of objects with \rm{CH$_{3}$OH} abundances (or with tighter upper limits) would be helpful to further constrain the \rm{CH$_{3}$OH} production mechanism(s).

\begin{figure}
\centering
\includegraphics[trim =0 0 0 0cm,width=0.45\textwidth]{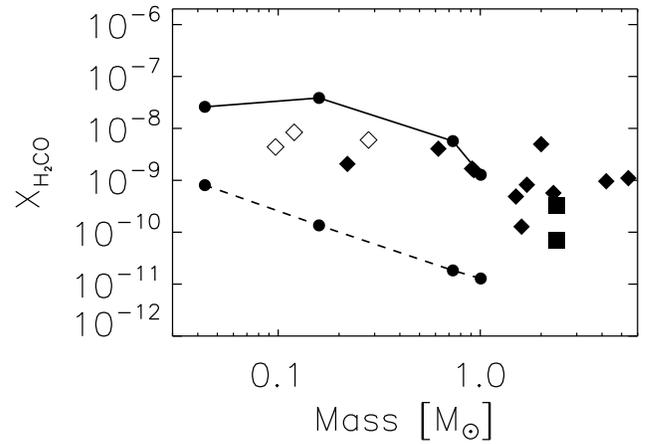} 
\caption{Global envelope abundances of \rm{H$_{2}$CO} versus envelope mass. Data for various objects from \cite{2005A&A...437..501J} are overplotted. Symbols and line styles are defined in Fig. \ref{XCOvsMenv}.
} 
\label{XH2COvsMenv}
\end{figure}

\begin{figure}
\centering
\includegraphics[trim =0 0 0 0cm,width=0.45\textwidth]{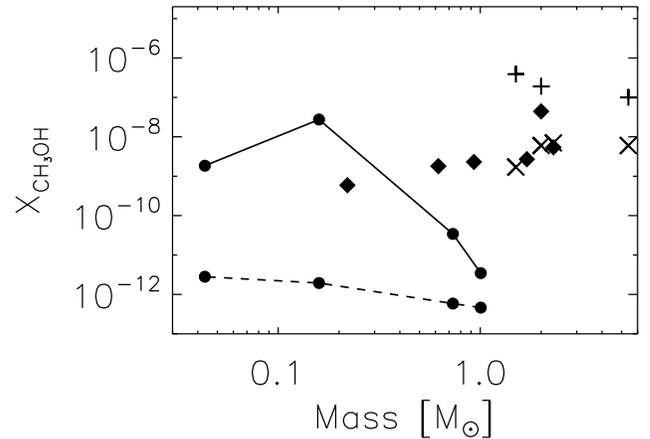} 
\caption{Global envelope abundances of \rm{CH$_{3}$OH} versus envelope mass. Data for various objects from \cite{2004A&A...416..603J,2002A&A...390.1001S} are overplotted. Symbols and line styles are defined in Fig. \ref{XCOvsMenv}. Other data points indicated are the abundances fitted with ``drop''-abundance profiles, with `$+$' the abundance in the inner region and `$\times$' that in the outer region of the envelope. } 
\label{XCH3OHvsMenv}
\end{figure}

\subsection{Ice composition}
The ice composition of grain mantles can be derived from observations of infrared absorption bands against a continuum. In dense molecular clouds ices are observed against background stars \citep[e.g.,][]{2001ApJ...558..185N, 2005ApJ...635L.145K, 2005ApJ...627L..33B,2007ApJ...655..332W}. For Class 0, Class I, and Class II objects the protostar itself provides the continuum against which we can observe ice absorption features. \rm{H$_{2}$O} ice is the most abundant ice-component followed by either \rm{CO} or \rm{CO$_{2}$}.  Other ices also exist but the abundances are generally lower. The \rm{H$_{2}$O} abundance on grains is a few times $10^{-4}$ and its abundance is relatively stable  because the binding energy is high (we have adopted $4674$ K). Therefore the abundances of the various ices are often expressed as percentage of \rm{H$_{2}$O}. In Model S the fractional  \rm{H$_{2}$O} ice abundance varies between $1.55$ $\times$ $10^{-4}$ and $1.80$ $\times$ $10^{-4}$ outside a radius of $\sim100$ AU, in agreement with observed values \citep[][and references therein]{2000ApJ...536..347G,1999A&A...343..966S}

The grain mantles in Class 0 objects have similar ice compositions as those in cold molecular clouds, although a small amount of thermal processing seems to have taken place for a number of sources, e.g., \cite{1999ApJ...522..357G, 2004ApJS..151...35G}. In Class I objects evidence is found for further processing of ices and complex molecule formation e.g, \cite{1998A&A...339L..17E, 2001A&A...376..254K, 2003A&A...401..613A, 2004ApJS..154..391W,2004ApJS..151...35G, 2004ApJS..154..359B,2007arXiv0712.2458Z}. 
Class II objects have heated the remaining envelope and outer layers of the disk above the sublimation temperature of ices. Although, in some edge-on objects ice are observed, probably arising from the cold midplane of the disk \citep[e.g.,][]{2005ApJ...622..463P}.

\begin{figure}
\centering
\includegraphics[trim =0.3cm 0 -0.3cm 0cm, width=0.5\textwidth]{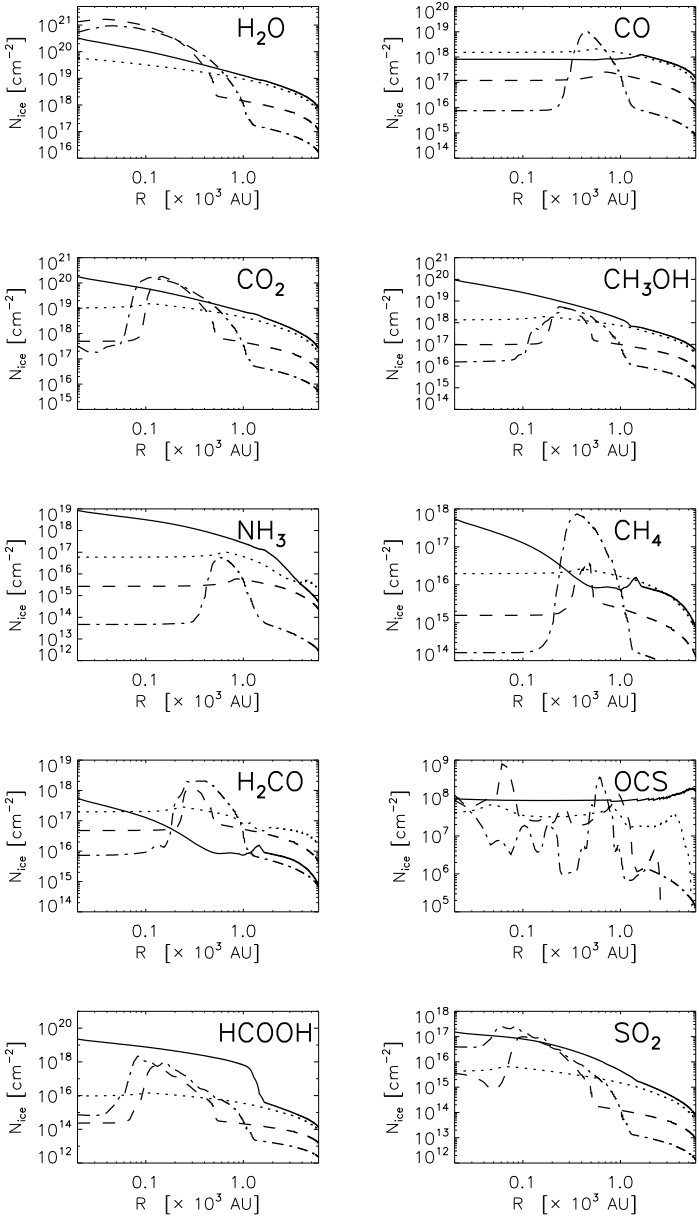} 
\caption{Column densities of various ices at four different times during the simulation,  $t = 0$ (solid lines), $t = 0.5$ $t_{\mathrm{ff}}$ (dotted), $t = 1.5$ $t_{\mathrm{ff}}$ (dashed), and $t = 2.5$ $t_{\mathrm{ff}}$ (dash-dotted).} 
\label{N_ice}
\end{figure}

The vertically integrated column densities for various ices (Model \rm{S}) are shown in Fig. \ref{N_ice}.
For most ices the overall evolutionary pattern is similar, we start with rising column densities towards the center at $t = 0$ in the prestellar core. Then a central sublimated zone develops where the column density remains relatively flat. The size of this region is determined by the binding energy of the species.  In the cold midplane of the disk, species with a low binding energy freeze-out again, while species with a high binding energy remain frozen out. Towards the disk, column densities increase by a few orders of magnitude for all ices. We will omit the results for Model \rm{G} as it fails to reproduce the observed ice-compositions for most of our surface species, see also Table \ref{tab:tabices}.

We compare the modeled ice fractions at four different times during the simulation ($t =$ 0, 0.5, 1.5, and 2.5 $t_{\mathrm{ff}}$) to observed values towards the field star \object{Elias 16} (in the vicinity of the dense core \object{TMC-1}),  the low-mass protostar \object{Elias 29}, and a sample of Class I/II objects from \cite{2007arXiv0712.2458Z}. The abundances are given relative to \rm{H$_{2}$O} and are averaged over the envelope (which does not include the disk). We have also included the ice composition within the disk, at a distance of $500$ AU, just outside the \rm{CO} sublimation front.  Model \rm{G} severely underproduces various ices like \rm{CH$_{3}$OH}, \rm{CO$_{2}$}, \rm{H$_{2}$O} and \rm{H$_{2}$CO}. This is explained by the fact that surface reactions are the dominant production mechanisms for these species. Sublimation of these ices then also leads to a higher gas phase abundance. Most of the \rm{CH$_{4}$} ice is located within the disk in Model \rm{S} where the density is high. For model \rm{G},  the \rm{CH$_{4}$} ice fraction is higher in the envelope. Surface reactions are thus not required to form a significant amount of  \rm{CH$_{4}$}. Below we further discuss the results from Model \rm{S} for several species.

Modeled \rm{CO} ice abundances are in rough agreement with observed values. Within the prestellar core the \rm{CO} ice fraction is on the low side with a value of 14\% while the \rm{CO$_{2}$} fraction of 47\% is quite high. Although, values of $37\%$ and $32\%$ were reported by \cite{2004ApJS..154..359B} towards \object{B5 IRS 1} and \object{HH 46 IRS}. The \rm{CH$_{3}$OH} ice fraction of about 5\% is in agreement with the observations given the large scatter, the same is true for \rm{H$_{2}$CO}. Our modeled \rm{SO$_{2}$} ice fraction ($\sim0.02\%$)  is on the low side, but its identification is uncertain. The \rm{NH$_{3}$} ice fraction ($\sim 0.5\%$) is low, observations indicate higher fractions of about $5\%$, but the \rm{NH$_{3}$} identifications are also somewhat uncertain. A higher binding energy as in \cite{2008ApJ...674..984A} would increase its fraction. As has been mentioned in Section \ref{sec:CN} the large uncertainties in the gas phase nitrogen-bearing chemistry may also have affected the \textrm{NH$_{3}$} ice abundance. During the collapse the fraction of \rm{CO} ice decreases from 15 to 5\%, the fraction of \rm{CO$_{2}$} ice also decreases, from 47 to 30\%. The \rm{CH$_{3}$OH} fraction stays more or less constant while the \rm{H$_{2}$CO} ice fraction increases from 1 to 4\%. The decrease in the \rm{CO} ice fraction is also being observed (as discussed above) and is the results of its low binding energy and the conversion into other more complex species. Slowing down the surface reaction $ \rm{CO} + \rm{O} \rightarrow \rm{CO_{2}}$ by increasing the activation energy of $1000$ K somewhat would produce less \rm{CO$_{2}$} and increase the \rm{CO} ice fraction. An alternative would be to increase the barrier against diffusion ($E_{\mathrm{B}}$) which should also slow down \rm{CO$_{2}$} production, as well as the production of other ices.
 
 \begin{table*}
\begin{center}
\caption{Grain mantle compositions}
\begin{tabular}{llllllllll}
\hline
\hline
Species$^{a}$& \object{Elias 16}$^{b}$& \object{Elias 29}$^{c}$ & Class \rm{I}/\rm{II} Objects$^{d}$ & Model &$t = 0.0$ &$t = 0.5$ $t_{\mathrm{ff}}$ &$t = 1.5$ $t_{\mathrm{ff}}$ &$t = 2.5$ $t_{\mathrm{ff}}$ & midplane disk$^{e}$\\
\hline
\rm{H$_{2}$O}$^{f}$
       & $100$ & 100 & 100&\rm{S} & $1.55(-4)$ & $1.65(-4)$ & $1.71(-4)$& $1.71(-4)$&$1.8(-4)$\\
       &              &         &    &\rm{G} &  {\sl 1.24(-5)}  & {\sl 8.29(-6)} & {\sl 4.5(-6)} &{\sl 4.2(-6)}  & {\sl 1.4(-5)}$^{g}$    \\
\rm{CO} & $25$ & 5.6 & \ldots &\rm{S} &13.9 & 17.0 & 11.6 & 5.0 & 17.6\\
               &            &        &   & \rm{G} &71.6&  61.1&  31.4 &13.5  &77.5\\
\rm{CO$_{2}$}      & $24$& 22 & $\sim 12$ &\rm{S}&46.9 &40.6&31.0&29.8&37.7\\
                                &           &       &                   &\rm{G}& 3.6(-1) & 3.1(-1)& 2.2(-1) &  2.2(-1) &4.2(-1)\\
\rm{CH$_{3}$OH}       & $< 2.3$ & $<4.4$ & $\gtrsim$ 2--9 &\rm{S}&8.7&7.0&6.6&6.7&2.2 \\
                                        &               &                &              &\rm{G} & 1.4(-5) &  2.1(-5) & 2.9(-5) & 2.9(-5) & 7.0(-8)   \\ 
\rm{NH$_{3}$} & $\leq 8$ & $< 9.2$ & $\lesssim 14$ &\rm{S}&$9.8(-1)$&$3.3(-1)$&$2.6(-1)$&$3.4(-2)$&$4.0(-1)$\\ 
                            &                &                &                           &\rm{G} & 1.7(-2) &   1.1(-2) &  2.7(-3) &  1.0(-3)  & 6.0(-3)           \\  
\rm{CH$_{4}$}       &  $<3$ & $< 1.6$ & $\sim 4$&\rm{S} & $1.1(-1)$&$1.7(-1)$&$1.4(-1)$&$1.0(-1)$&1.8\\
                                  &           &               &                  &\rm{G} &  2.2 &    2.0&    1.6 &  1.4  & 1.0 \\ 
\rm{H$_{2}$CO}& \ldots & $<1.8$& $\sim 2$&\rm{S}&1.1&1.6&3.9&4.1&2.9\\
                             &             &             &                 &\rm{G} &  4.5(-4)  &   3.7(-4) &  4.3(-4) & 4.7(-4) &4.8(-4)  \\
\rm{OCS}&  $< 0.2$ & $< 0.05$&\ldots&\rm{S}&$4.3(-9)$&$4.9(-10)$&$4.6(-10)$&$8.1(-10)$&$0$\\
                &                 &                  &   &    \rm{G}  &3.7(-3)    &       4.0(-3) &    3.0(-3) &    2.7(-3)  &9.1(-2)       \\
\rm{HCOOH}&  \ldots& $< 0.9$& $\sim 0.6$&\rm{S}&$1.5$&$2.6(-2)$&$1.4(-2)$&$1.4(-2)$&$3.4(-2)$\\
                     &      &              &                   &  \rm{G} & 7.7(-3)  &  9.3(-3) &    7.7(-3) &  7.1(-3) &  6.9(-1) \\
\rm{SO$_{2}$}& \ldots& \ldots& $\sim 0.5$&\rm{S}&$2.3(-2)$&$1.3(-2)$&$8.7(-3)$&$8.4(-3)$&$1.2(-2)$\\
                          &     &   &                    &\rm{G} &      7.8(-4)   &    1.0(-3)  &    7.0(-4)   &   5.9(-4)  &   8.3(-3)  \\
\hline
\end{tabular}
 \label{tab:tabices}
\end{center}
$^{a}$ abundances are given as $a(b) = a \times 10^{b}$ 

$^{b}$ \rm{CO} \cite{1995ApJ...455..234C}; \rm{CO$_{2}$}, \rm{CH$_{3}$OH}, \rm{NH$_{3}$}, and \rm{CH$_{4}$} \cite{2005ApJ...635L.145K}; \rm{OCS} \cite{1997ApJ...479..839P}  

$^{c}$ \cite{2000IAUS..197..135E} and \cite{2004ASPC..309..547B} 

$^{d}$ \cite{2007arXiv0712.2458Z}

$^{e}$ at $t = 2.5$ $t_{\mathrm{ff}}$ and a radius of $\sim500$ AU

$^{f}$ \rm{H$_{2}$O} fractional abundance are given for the simulation: $n(\mathrm{H}_{2}\mathrm{O})/n(\mathrm{H}_2)$

$^{g}$ for model \rm{G} we have taken an \rm{H$_{2}$O} fractional abundance of $1.8 \times 10^{-4}$ for calculating the ice fraction of other species, instead of the reported values on this line\vspace{2mm}

Observed ice compositions as percentage of the \rm{H$_{2}$O} abundance compared with model results. Every first line for a species shows the results from Model \rm{S} (includes surface reactions) and the second line shows the results for Model \rm{G} (excludes surface reactions)

\end{table*}

\section{Discussion}
\label{sec:discussion}
When we compare our gas phase abundances and ice compositions to observations, it seems that the surface reaction rates are somewhat too high. First we slightly overproduce the amount of  \rm{CO$_{2}$} compared to \rm{CO}. Our ratio of \rm{CO$_{2}$} to \rm{CO} is about three while observations often show ratios of about one to two. Also the \rm{H$_{2}$CO} gas phase abundances are a little high. Furthermore species like \rm{HCN}, \rm{HNC} are overproduced by quite some amount, although uncertainties in the nitrogen-bearing chemistry might also have influenced these results. To slow down the surface rates several options exist. First, the binding energy ($E_{\mathrm{b}}$) of atomic hydrogen could be increased which would slow down surface hydrogenation reactions. We have taken a fast diffusion rate for atomic hydrogen ($E_{\mathrm{b}} = 492$) so it could be that we have to increase its binding energy to slow down the diffusion rate \citep[see][]{2000MNRAS.319..837R}. Another option is to increase the barrier against diffusion $E_{\mathrm{B}}$ for all surface reactions (e.g., $E_{\mathrm{B}} = 0.5E_{\mathrm{b}}$).
For certain reactions it is also possible to increase the activation energy ($E_{a}$). Although, this will not help slowing down the \rm{HCN{ and \rm{HNC} production as there are are no activation energies involved in the production of these species but for the oxidation reaction of \rm{CO} leading to \rm{CO$_{2}$} this would be an option. As a first step, lowering the diffusion rate for atomic hydrogen and increasing the energy barrier against diffusion will be the best option. 

The underproduction of \rm{CS} and \rm{SO} (Model S) is not completely understood. Although \rm{CS} and \rm{SO} are transformed by surface reactions this should not have a very large effect on the abundances as most of those surface reactions do not involve \rm{H} atoms and are therefore quite slow. Reduced gas phase abundances of \rm{CO} (and other species) could result in a different gas phase chemistry leading to the decrease of parent species for \rm{CS} and \rm{SO} production. More work is needed to further investigate this effect.

In our simulation we have taken a constant value for the extinction ($A_{\mathrm{V}} = 15$) throughout the collapsing prestellar core. This is a good approximation during the initial phases of the collapse ($t_{\mathrm{ff}}$  $\lesssim 1$). 
Only at the outer edges of the core the results could change significantly due to the presence of the interstellar radiation field combined with lower values for the extinction. We are mainly interested in the central regions of the core and have only plotted our results for $r \lesssim 5000$ AU. Prestellar cores can also be shielded by several magnitudes of external extinction, for example if they are embedded within a larger dark molecular cloud. Our simulation can therefore be described as the chemical modeling of an externally shielded collapsing core. At $t_{\mathrm{ff}}$  $\gtrsim 1$ the extinction in the region above the midplane of the disk drops below $A_{\mathrm{V}} < 5$ so that photo-dissociation and ion-molecule reactions become important. The importance of these reactions is increased by the stronger radiation field from the central protostar. However, the overall effect on the global abundances of species is reduced as most of the mass is located within the shielded disk where the extinction is large ($A_{\mathrm{V}} > 15$), even at the closest distance to the protostar of $\sim7$ AU within the simulation. 

To include the effect of the radiation field the extinction should be calculated at every position within the core. From the density given by the hydrodynamical simulation we could in principle derive the amount of extinction using the relation $A_{\mathrm{V}} =  \frac{N_{\mathrm{H}}}{1.59 \times 10^{21}}$ {\tiny  $\frac{\mbox{mag}}{\mbox{cm}^{-2}}$  }, with $N_{\mathrm{H}}$ the column density of hydrogen nuclei between the protostar and the point of interest.  However, in practice this is more complicated as a significant amount of extinction arises from regions close to the protostar within $7$ AU, beyond the resolution of our simulation where we do not have information on the density distribution. The dust destruction radius is located roughly at $2000$ K \citep{1996A&A...312..624D} $\sim0.1$ AU from the protostar which sets an upper limit on the resolution needed to derive the amount of extinction. By increasing the resolution of the hydrodynamical simulation this can be satisfied. As a first approximation for the intrinsic strength of radiation field, we could take the luminosity from the protostar given by the hydrodynamical simulation. The protostar luminosity is based on the mass accretion rate plus a central core luminosity. Including these effects can be a starting point for future work. 
Our hydrodynamical simulation has a simplified radiative transfer, this may have resulted in deviations in the gas and dust temperature compared to more realistic simulations. \cite{2004A&A...428..511J} presented a model which calculated the gas and dust temperature as well as the chemistry in the surface layers of a flaring protoplanetary disk. The heating rate of the gas was controlled by UV radiation and the gas cooling by line emission of various species and by collisions with dust grain. They found that the gas temperature exceeds the dust temperature, in the part of the disk that is optically thin to UV radiation, by up to several hundreds of Kelvin. In the optically thick part of the disk the gas temperature was found to be well coupled to the dust temperature. In our hydrodynamical simulation the dust isotherms are more spherical compared to the results in \cite{2004A&A...428..511J}. Small changes in the dust temperature can have an important effect on the sublimation rate as it is exponentially dependent on the dust temperature. Therefore accurate dust temperatures are important for calculating the position of the sublimation front if the binding energies of the species are known. Observations of the position of sublimation fronts of various species could be a sensitive tracer of the dust temperature. The very high gas temperatures found above the disk also have an effect on the gas phase chemistry. Although  \cite{2004A&A...428..511J} reported that effect was small. However, their chemical network was rather limited and no gas-grain interactions were included. A temperature increase from $60$ to $550$ K in the upper disk layers (at $R = 150$ AU and $z\sim90$ AU) will increase the freeze-out rate by a factor of $\sim 3$, although this may be compensated by a lower sticking coefficient $S$. X-rays from the central protostar also have an effect on the chemistry as shown by  \cite{2005A&A...440..949S}. Abundances of some species were found to be enhanced by $2-3$ orders of magnitude. Due to smaller cross sections, X-rays penetrate further into the disk and envelope and therefore affect the chemistry on larger scales ($\gtrsim 1000$ AU) than the UV radiation field ($\sim400$ AU, \cite{2004A&A...425..577S}).  Eventually the chemistry needs to be coupled to the radiative transfer \citep[e.g., ][]{2003A&A...397..789V} for the modeling to be self consistent and correctly address this issues.

In our model we can follow the chemical composition of gas accreting onto the disk. Gas does not directly accrete onto the disk but flows around the disk upwards (see Fig. \ref{semenov}) before entering the disk around the centrifugal radius (see also Paper \rm{I}). This has important implications for the chemistry as the gas is more exposed to UV and X-ray radiation from the central protostar before entering the disk. The dust temperature at the centrifugal radius of the disk sets a sublimation temperature for ices accreting onto the disk. The result is that ices with a low binding energy  ($E_{\mathrm{b}} \lesssim1200$ K) sublimate before entering the disk while ices with a high binding energy ($E_{\mathrm{b}} \gtrsim1200$ K) are able to enter the disk. The strong UV radiation field may also cause photodesorption of ices. Recently, \cite{2007ApJ...662L..23O} determined high yields for \rm{CO} photodesorption in laboratory experiments. In our simulation at $t = 2.5$ $t_{\mathrm{ff}}$, most of the \rm{CO} ice ($\sim75\%$) is not able to accrete onto the disk, as the sublimation front lies just outside of the centrifugal radius (at $\sim500$ AU), while \rm{H$_{2}$CO} ice can enter the disk without sublimating, see Fig. \ref{semenov}. By following the chemical evolution of gas accreting onto the disk we can derive the initial abundances to be adopted in more detailed modeling of the chemistry within the disk. For example \cite{2004A&A...417...93S} took as initial abundances those computed in a dark cloud (out of which the disk eventually forms). However, the initial abundances are different from those in dark clouds as some chemical processing has already taken place before the gas enters the disk. Using the correct initial abundances might improve the results of those models.

\begin{figure}
\centering
\includegraphics[trim =0.5cm 0cm -0.5cm 0cm,width=0.5\textwidth]{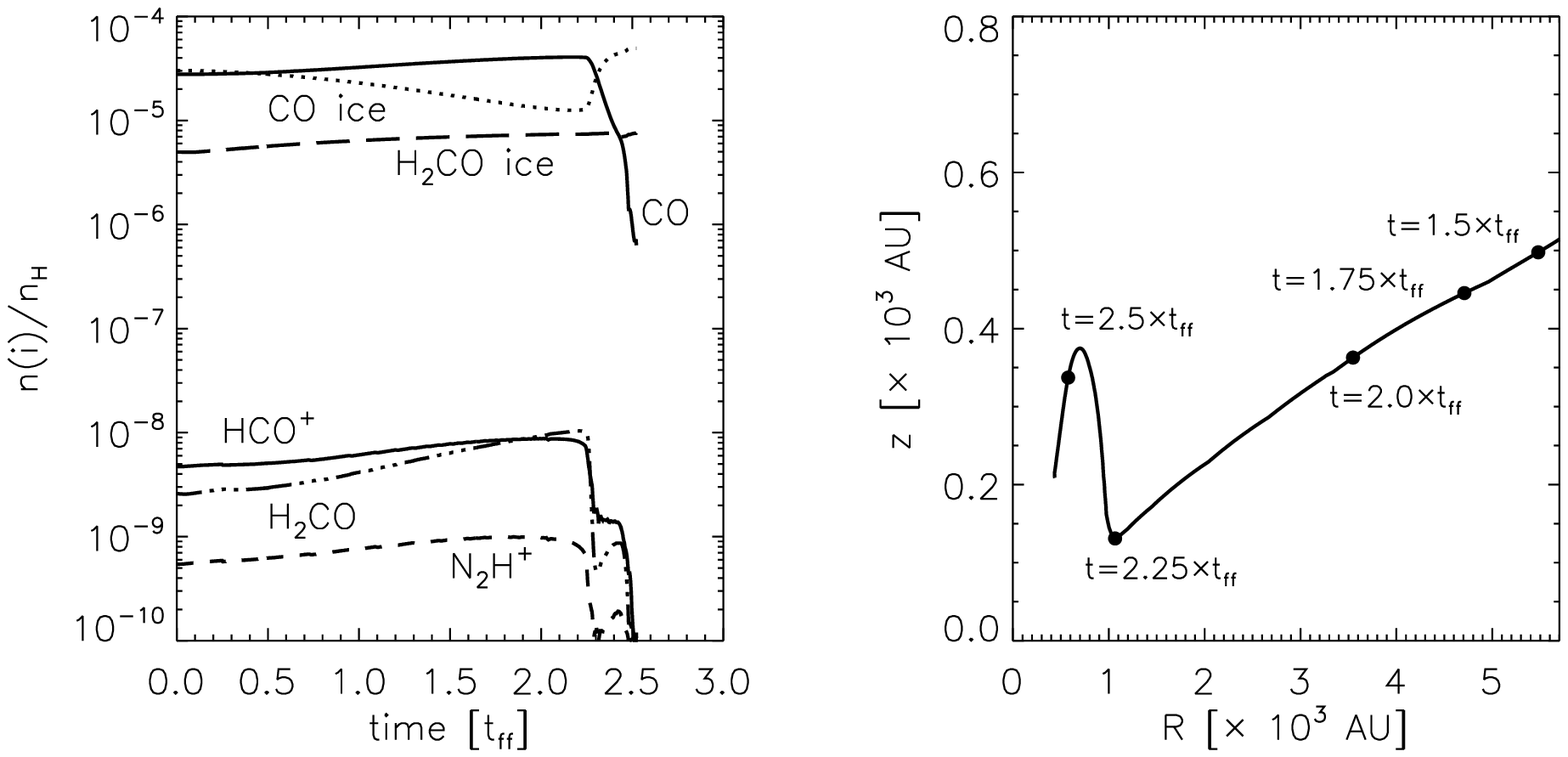} 
\caption{Left: Abundances as function of time for a single trace particle which accretes onto the disk. At $t = 1.5 t_{\mathrm{ff}}$  the conditions for the trace particle are roughly that of the dark ISM, $T = 10$ K and $n = 10^{4}$ cm$^{-3}$.  Right: Trace particle trajectory, the position of the particle at different times is indicated in the figure. The acceleration of the particle as well as the upward motion around the disk are visible before the particle accretes onto the disk.
} 
\label{semenov}
\end{figure}

We have chosen a standard dust grain radius of 0.1  $\mu$m for our simulation to compare our results with previous work \citep[e.g.,][]{2008ApJ...674..984A}. This is an oversimplification as grains are expected to grow to larger sizes under the conditions in the simulation, especially within the disk. The main effect of grain growth will be a decrease in the depletion rate as the total grain surface area decreases. The lower depletion rates also have an effect on the nitrogen-bearing chemistry as was shown by \cite{2006A&A...456..215F}. An approach might be to use the more sophisticated treatment by \cite{2005A&A...436..933F}, following the coagulation of the grains simultaneously with the chemical evolution of the medium. This is, however, beyond the scope of this paper.

\section{ Conclusions}
\label{sec:conclusion}
We have modeled the chemical abundances during prestellar core collapse and have compared  the global evolution of these abundances with observations.  We have used trace particles, moving with the flow of the gas as given by a 2-dimensional hydrodynamics simulation, to derive the chemical evolution of the collapsing core.

Gas-grain interactions determine to a large extent the abundance evolution of species during prestellar core collapse. Before collapse, most species are depleted due to freeze-out onto dust grains. When the temperature rises in the center a sublimated zone develops. At the same time in the outer parts of the envelope the density drops, which lowers the freeze-out rate resulting in an increase in the gas phase abundance of species with a relatively low binding energy. Within the disk, species with a low binding energy freeze-out again, species with a higher binding energy are able to enter the disk without sublimating. The abundances of charged species are mostly determined by the balance between destruction and formation. Parent species, or destructors often have gas-grain interactions so the abundance of charged species is still influenced by gas-grain interactions, although in an indirect way. 

We conclude that grain surface reactions are important for the formation of some species, in particular for \rm{CH$_{3}$OH}, \rm{CO$_{2}$}, \rm{H$_{2}$O} and \rm{H$_{2}$CO}. Without surface reactions most of the ice is in the form of \rm{CO}, and the relative fraction of other ices is much lower than observed. An example is \rm{CH$_{3}$OH} where the ice fraction for Model \rm{G} is lower by more than 5 orders of magnitude. Gas phase abundances of species likes \rm{CH$_{3}$OH} and \rm{H$_{2}$CO} are also too low without surface reactions when we compare global envelope abundances with observations. So we confirm that surface reactions are important for the production of some species.
\rm{CH$_{4}$} is produced quite efficiently in the gas phase. Only at high densities within the disk \rm{CH$_{4}$} is produced more efficiently by grain surface reactions. Abundances are higher in the envelope if we do not include surface reactions. It could be that some \rm{CH$_{4}$} is transformed in Model \rm{S} towards \rm{CH$_{3}$} and \rm{C$_{2}$H$_{2}$} by surface reactions.  \rm{CS} and \rm{SO} abundances show a better match to observations if no surface reactions are included.

Our method provides important information on the initial abundances of species to be adopted in chemical modeling of protoplanetary disks. The initial abundances are often derived from chemical models of dark molecular clouds. However, as we have shown, significant chemical processing of the gas and the grain mantles occurs before accretion onto the disk. For example ices with a low binding energy ($\lesssim 1200$ K) are not able to enter the disk, while those with a higher binding energy are able to directly accrete onto the disk. Also gas accreting onto the disk (coming from $z \lesssim 1000$ AU) may be temporally exposed to the strong radiation field from the central protostar, especially at the later stages of the collapse when $t \gtrsim 2$ $t_{\mathrm{ff}}$. This is caused by the upward motion of gas around the disk, representing a bow shock feature. The gas then accretes above the midplane, roughly at the position of the centrifugal radius onto the disk.

The method we have presented for modeling the chemical abundances is generally applicable to other hydrodynamical simulations. Although, modeling the chemical evolution is computationally expensive it can be parallelized easily as the trace particles have no interaction with each other. This can be used to extend the modeling to higher spatial resolution (from the $20$ AU we have used) and towards three dimensions. This will allow us to probe the chemical evolution in the planet forming zone in greater detail. Including the effects of the UV radiation and X-rays is important in the later stages ($t_{\mathrm{ff}} \lesssim 1$) of the collapse, especially for the upper disk layers and the envelope.

Finally, modeling line profiles with RATRAN and comparing them to observations may provide more information on the physical conditions during prestellar core collapse.

From an observational point of view, increasing the sample of objects with modeled envelope masses and abundances, will help to test chemical models. In particular, more \rm{CH$_{3}$OH} observations as well as those from sulfur bearing species are needed. This will also help to find species which could be used as chemical clocks. Second, laboratory experiments of grain surface reactions may further help to constrain the surface reaction and diffusion rates.

\begin{acknowledgements}
      We would like to thank the anonymous referee for useful comments. CB is partially supported by the European Commission through the FP6 - Marie Curie Early Stage Researcher Training programme. The research of MRH is supported through a VIDI grant from the Netherlands Organization fro Scientific Research.
    
\end{acknowledgements}

\bibliographystyle{aa}
\bibliography{references}
\newpage
 \Online

 \begin{appendix} 
 
 \section{2-Dimensional abundances including grain surface chemistry}
\label{appendix1}

 \begin{figure}[h!]
 \centering
 \includegraphics[trim =0 0 0 0cm,width= 0.50\textwidth]{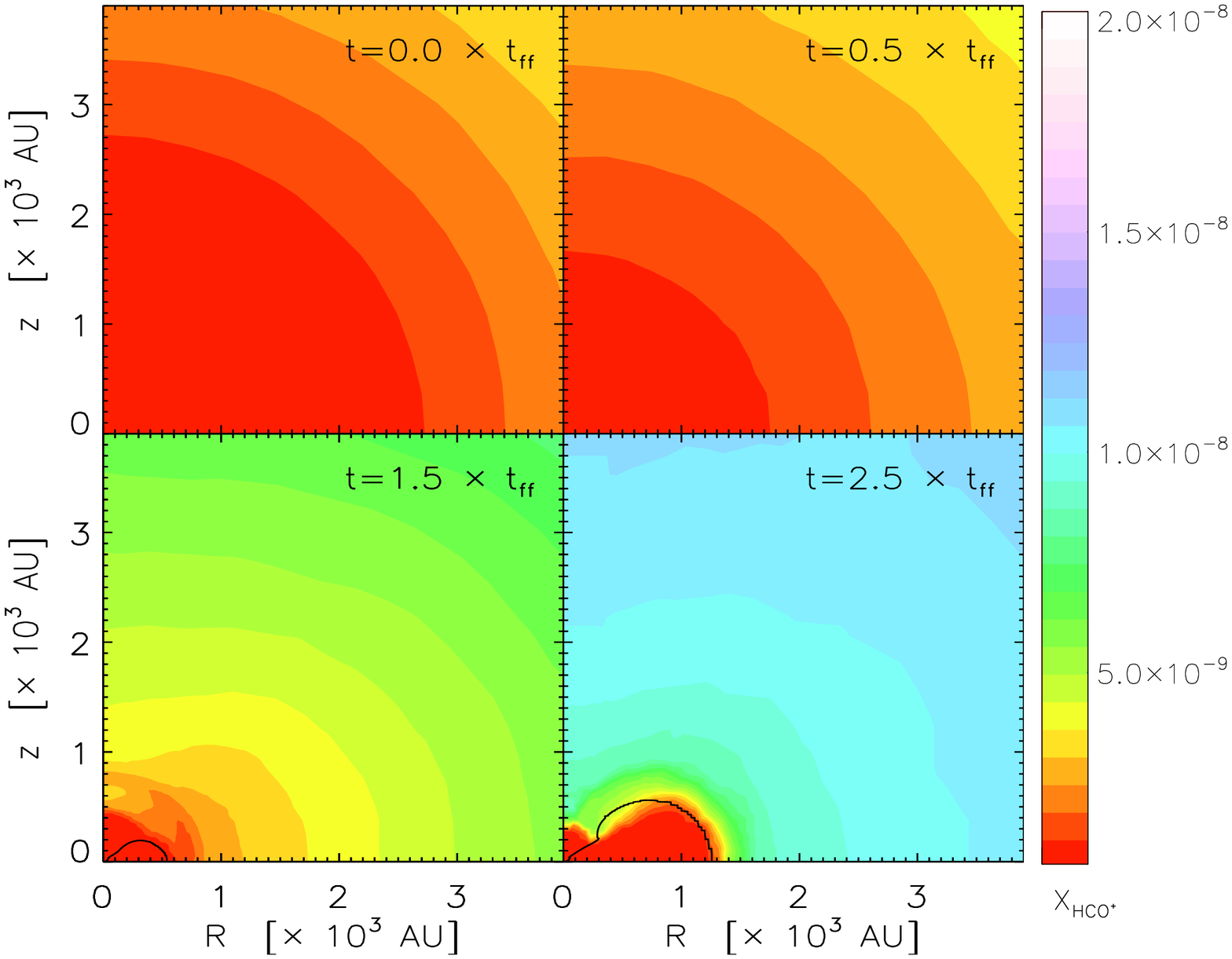}
 \caption{Distribution of \rm{HCO$^{+}$} throughout the core at four different times during the simulation for Model \rm{S}. The black contour shows the outline of the disk. The disk is defined as the region where $\sqrt{v_z^2 + v_R^2} \leq v_\phi$, with $v_\phi$ the rotational velocity.}
 \label{appfig:HCO+}
 \end{figure}

 \begin{figure}[h!]
 \centering
 \includegraphics[trim =0 0 0 0cm,width= 0.50\textwidth]{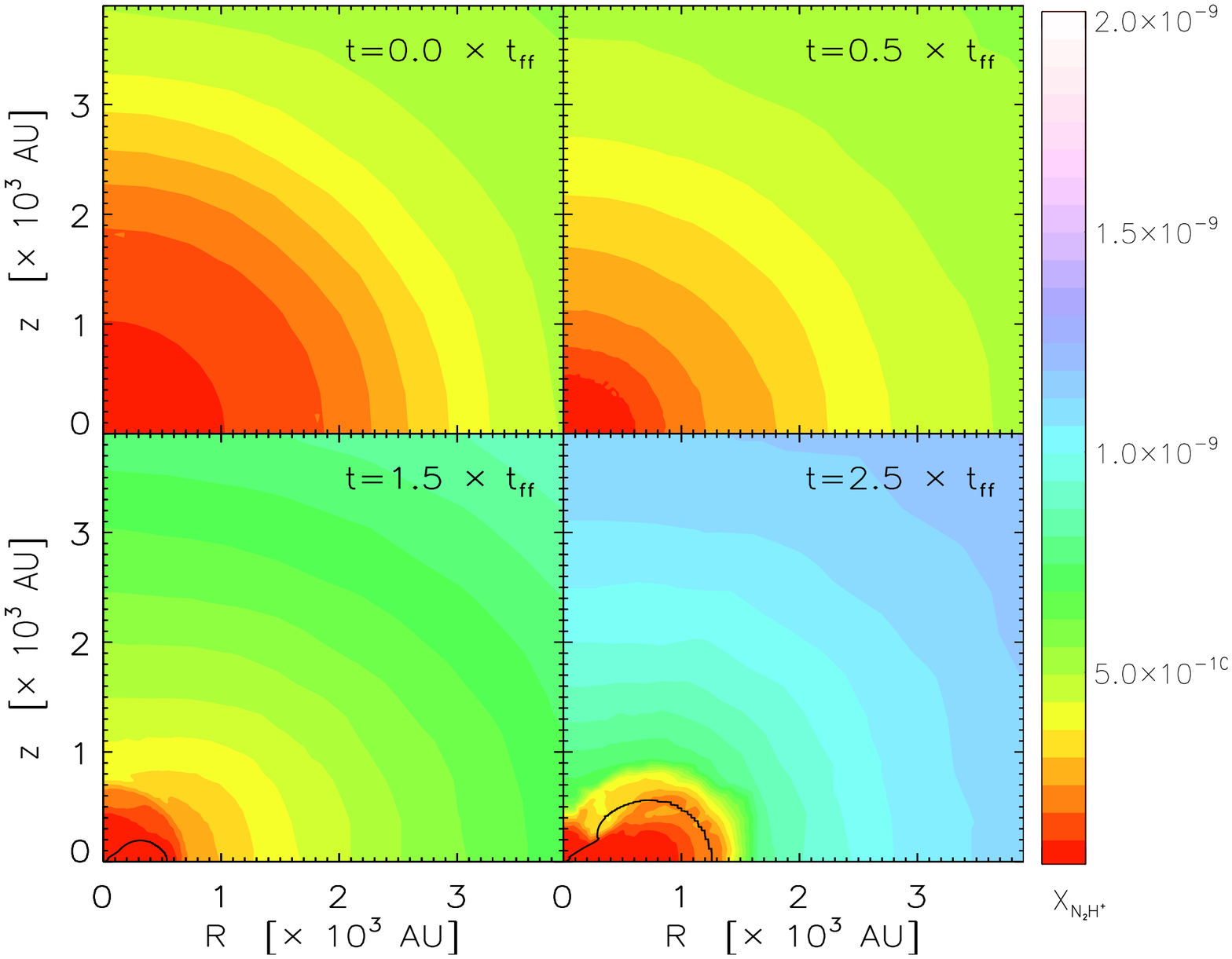}
 \caption{Same as for Fig. \ref{appfig:HCO+} but for \rm{N$_{2}$H$^{+}$}}
 \label{appfig:N2H+}
 \end{figure}
 
  \begin{figure}[h!]
 \centering
 \includegraphics[trim =0 0 0 0cm,width= 0.50\textwidth]{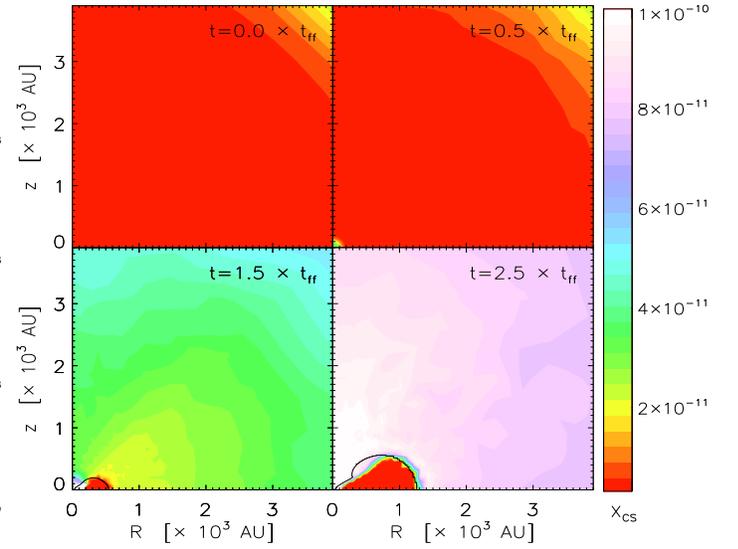}
 \caption{Same as for Fig. \ref{appfig:HCO+} but for \rm{CS}}
 \label{appfig:CS}
 \end{figure}

 \begin{figure}[h!]
 \centering
 \includegraphics[trim =0 0 0 0cm,width= 0.50\textwidth]{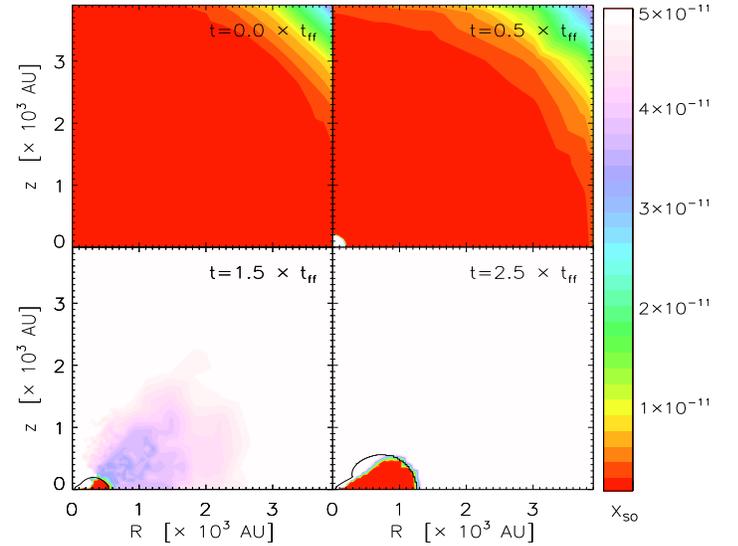}
 \caption{Same as for Fig. \ref{appfig:HCO+} but for \rm{SO}}
 \label{appfig:SO}
 \end{figure}
 
  \begin{figure}[h!]
 \centering
 \includegraphics[trim =0 0 0 0cm,width= 0.50\textwidth]{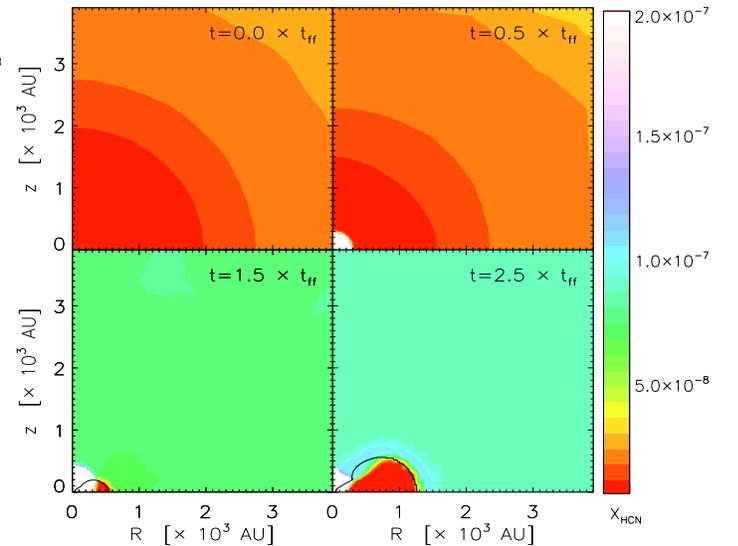}
 \caption{Same as for Fig. \ref{appfig:HCO+} but for \rm{HCN}}
 \label{appfig:HCN}
 \end{figure}

  \begin{figure}[h!]
 \centering
 \includegraphics[trim =0 0 0 0cm,width= 0.50\textwidth]{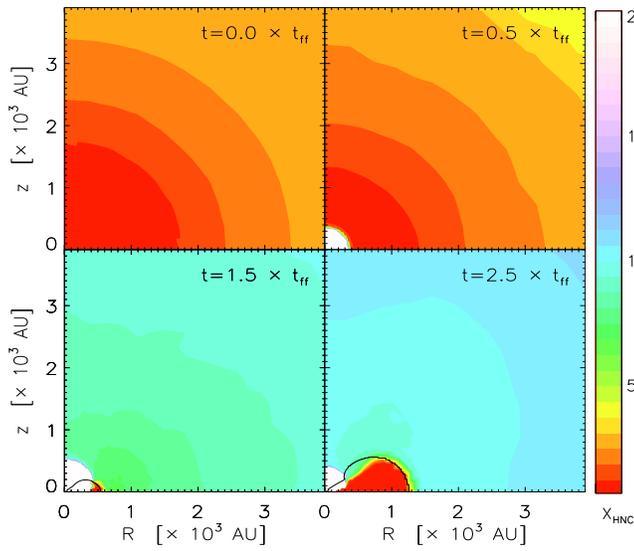}
 \caption{Same as for Fig. \ref{appfig:HCO+} but for \rm{HNC}}
 \label{appfig:HNC}
 \end{figure}
 
   \begin{figure}[h!]
 \centering
 \includegraphics[trim =0 0 0 0cm,width= 0.50\textwidth]{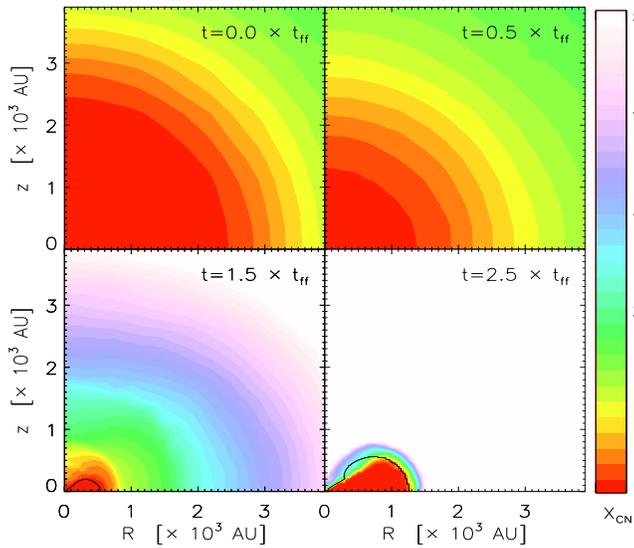}
 \caption{Same as for Fig. \ref{appfig:HCO+} but for \rm{CN}}
 \label{appfig:CN}
 \end{figure}
 
    \begin{figure}[h!]
 \centering
 \includegraphics[trim =0 0 0 0cm,width= 0.50\textwidth]{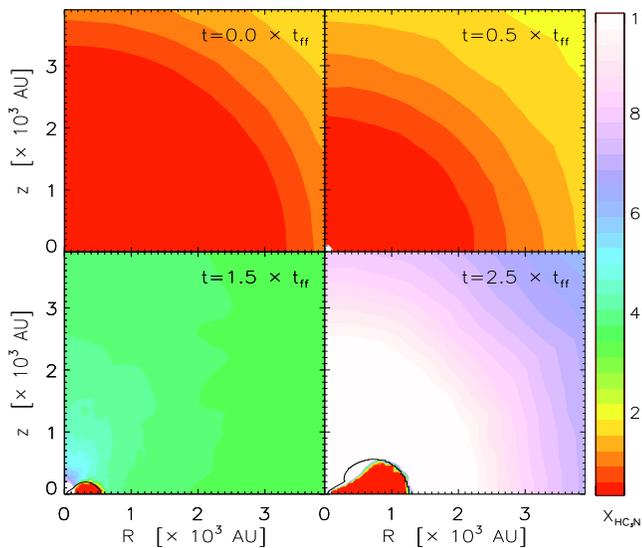}
 \caption{Same as for Fig. \ref{appfig:HCO+} but for \rm{HC$_{3}$N}}
 \label{appfig:HC3N}
 \end{figure}
 
     \begin{figure}[h!]
 \centering
 \includegraphics[trim =0 0 0 0cm,width= 0.50\textwidth]{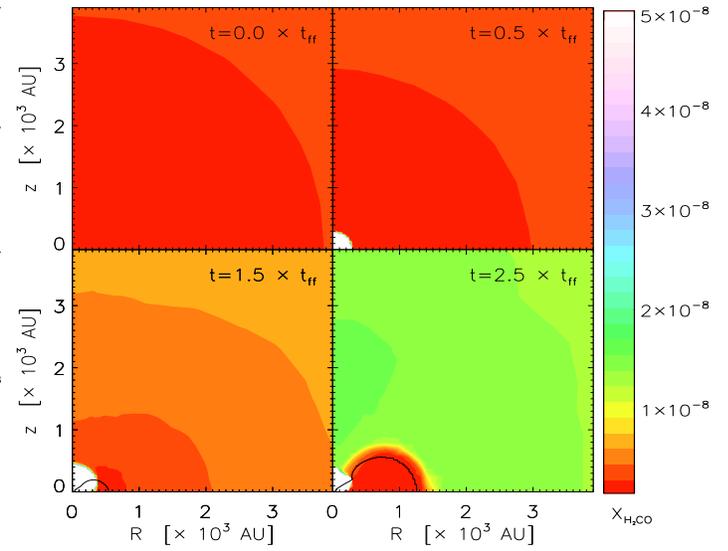}
 \caption{Same as for Fig. \ref{appfig:HCO+} but for \rm{H$_{2}$CO}}
 \label{appfig:H2CO}
 \end{figure}
 
     \begin{figure}[h!]
 \centering
 \includegraphics[trim =0 0 0 0cm,width= 0.50\textwidth]{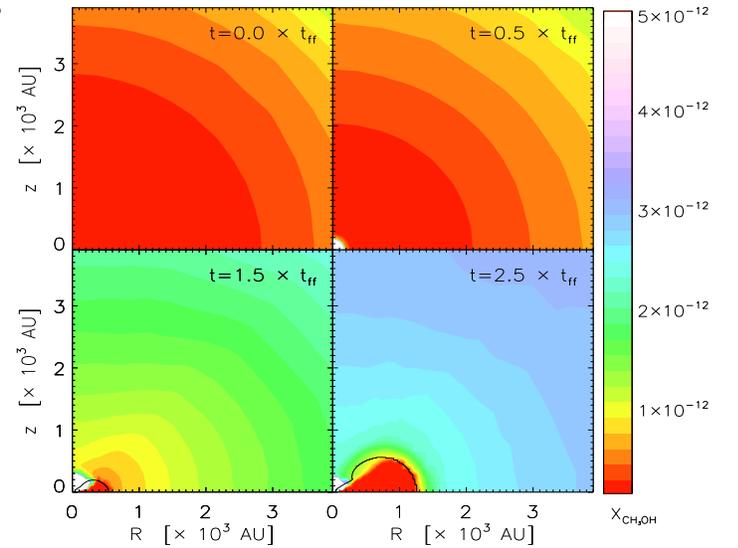}
 \caption{Same as for Fig. \ref{appfig:HCO+} but for \rm{CH$_{3}$OH}}
 \label{appfig:CH3OH}
 \end{figure}
\clearpage 
 \section{2-Dimensional abundances without grain surface chemistry}
 \label{appendix2}
 \begin{figure}[h!]
 \centering
 \includegraphics[trim =0 0 0 0cm,width= 0.50\textwidth]{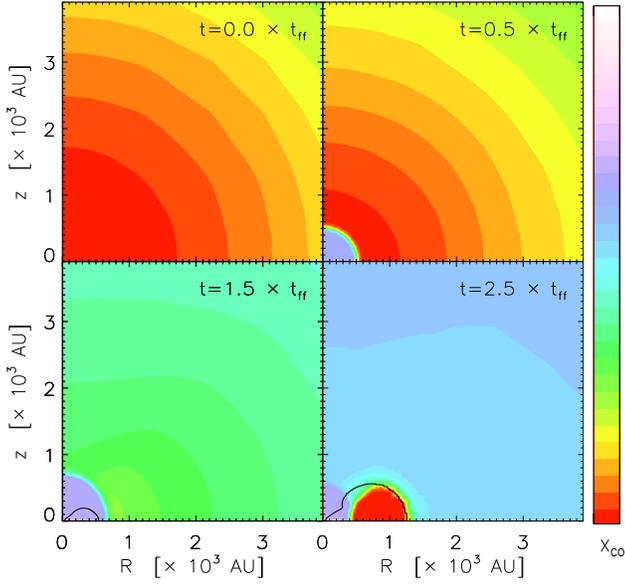}
 \caption{Distribution of CO throughout the core at four different times during the simulation for Model \rm{G}.}
 \label{appfig:CO_noschem}
 \end{figure}

 \begin{figure}[h!]
 \centering
 \includegraphics[trim =0 0 0 0cm,width= 0.50\textwidth]{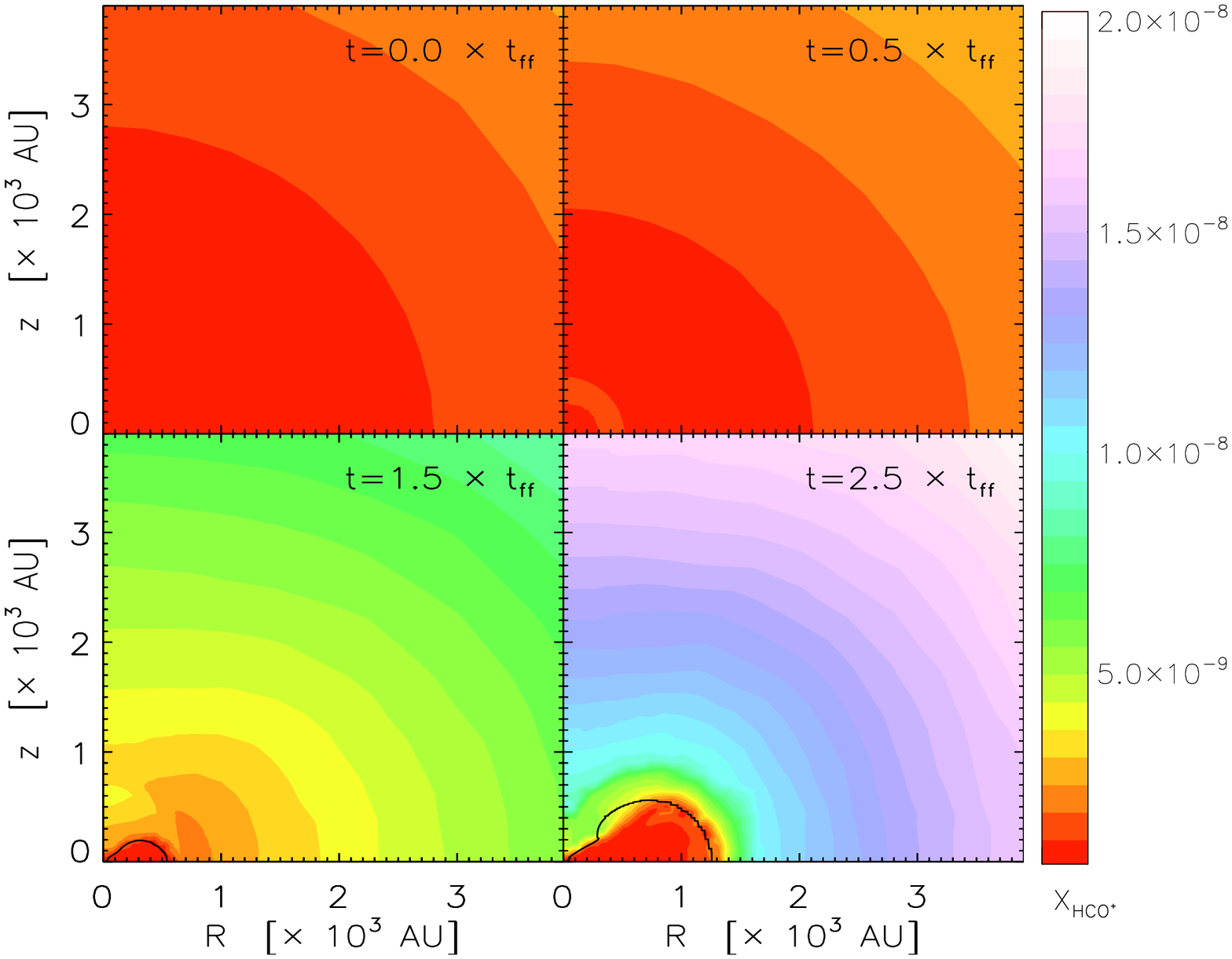}
 \caption{Same as for Fig. \ref{appfig:CO_noschem} but for \rm{HCO$^{+}$}}
 \label{appfig:HCO+_noschem}
 \end{figure}

 \begin{figure}[h!]
 \centering
 \includegraphics[trim =0 0 0 0cm,width= 0.50\textwidth]{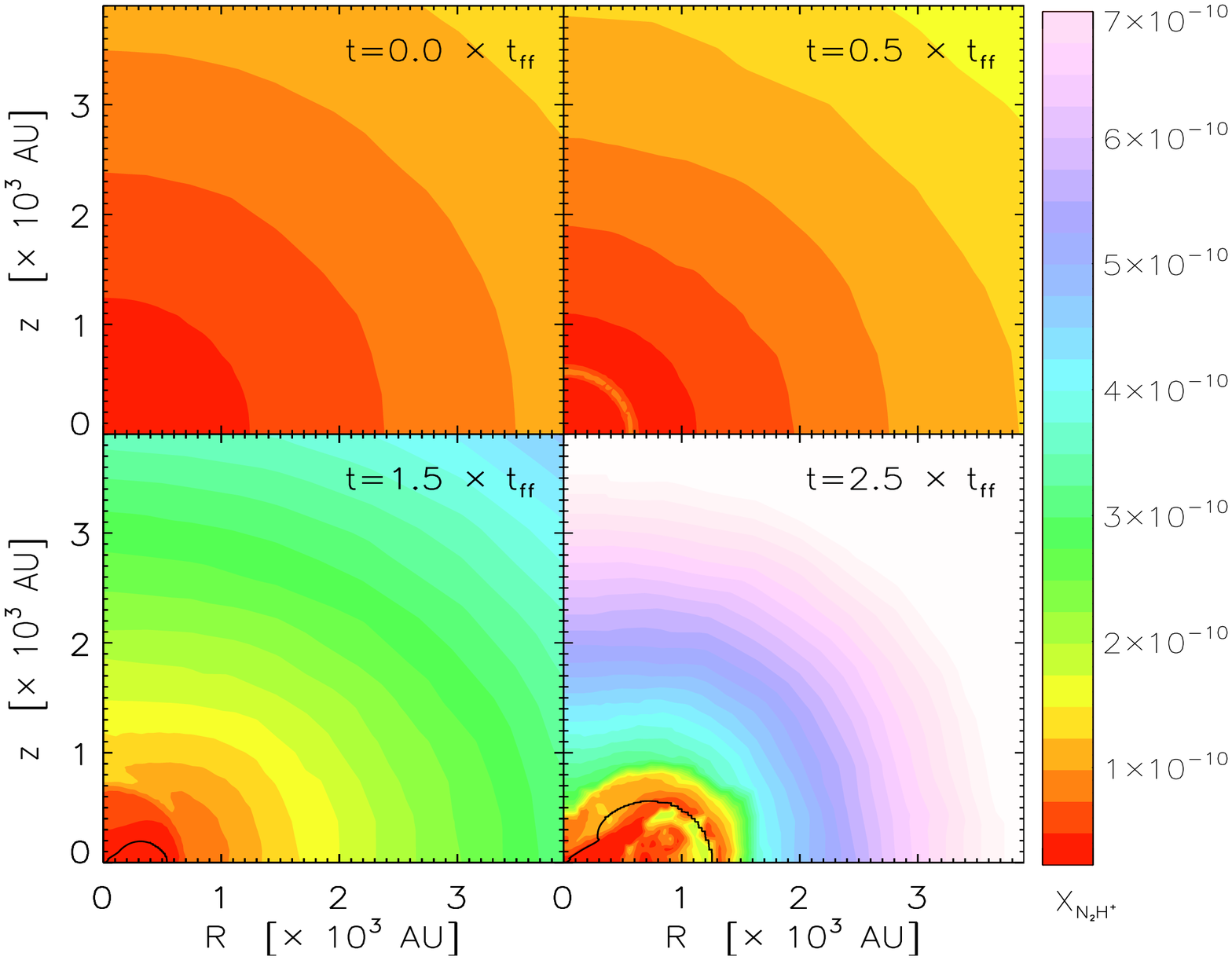}
 \caption{Same as for Fig. \ref{appfig:CO_noschem} but for \rm{N$_{2}$H$^{+}$}}
 \label{appfig:N2H+_noschem}
 \end{figure}
 
  \begin{figure}
 \centering
 \includegraphics[trim =0 0 0 0cm,width= 0.50\textwidth]{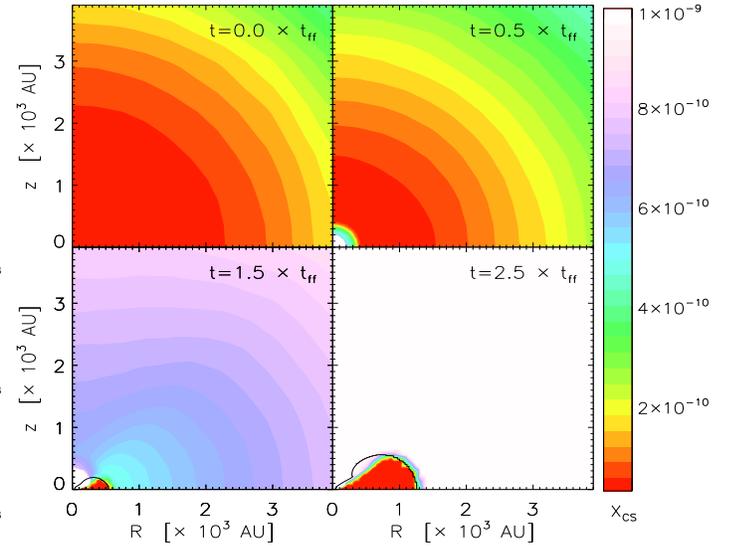}
 \caption{Same as for Fig. \ref{appfig:CO_noschem} but for \rm{CS}}
 \label{appfig:CS_noschem}
 \end{figure}

 \begin{figure}
 \centering
 \includegraphics[trim =0 0 0 0cm,width= 0.50\textwidth]{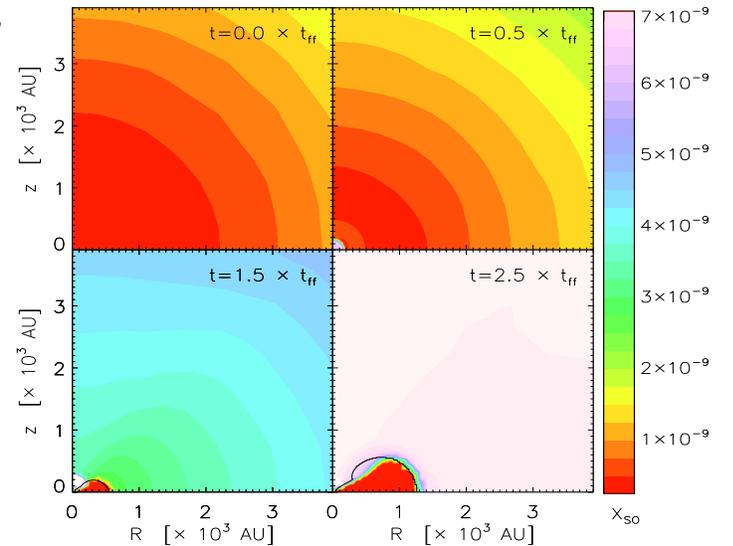}
 \caption{Same as for Fig. \ref{appfig:CO_noschem} but for \rm{SO}}
 \label{appfig:SO_noschem}
 \end{figure}
 
  \begin{figure}
 \centering
 \includegraphics[trim =0 0 0 0cm,width= 0.50\textwidth]{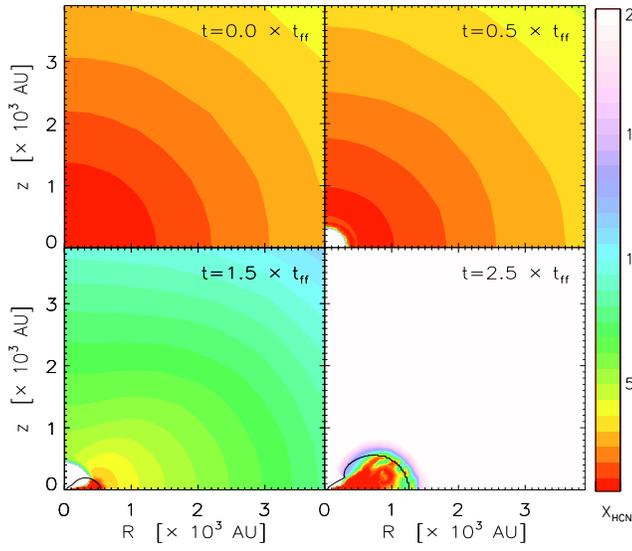}
 \caption{Same as for Fig. \ref{appfig:CO_noschem} but for \rm{HCN}}
 \label{appfig:HCN_noschem}
 \end{figure}

  \begin{figure}
 \centering
 \includegraphics[trim =0 0 0 0cm,width= 0.50\textwidth]{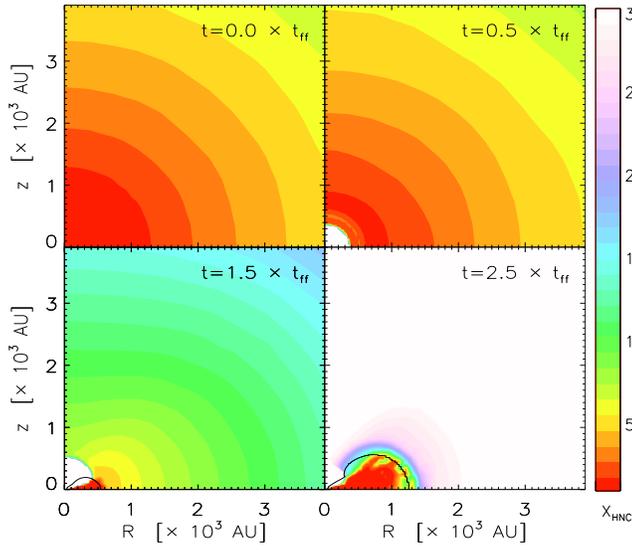}
 \caption{Same as for Fig. \ref{appfig:CO_noschem} but for \rm{HNC}}
 \label{appfig:HNC_noschem}
 \end{figure}
 
   \begin{figure}
 \centering
 \includegraphics[trim =0 0 0 0cm,width= 0.50\textwidth]{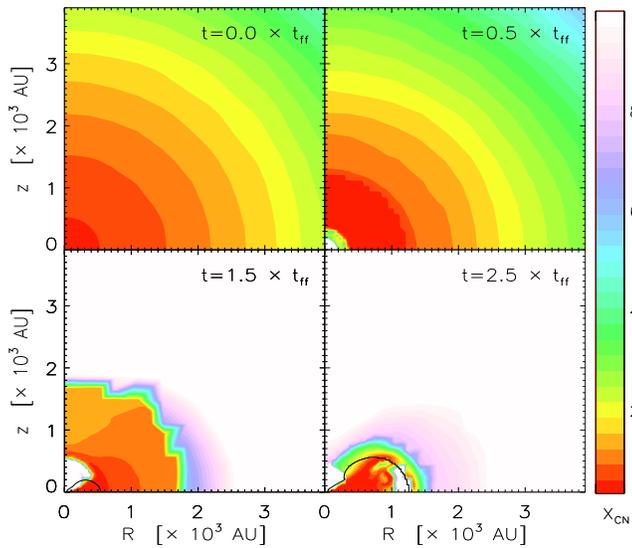}
 \caption{Same as for Fig. \ref{appfig:CO_noschem} but for \rm{CN}}
 \label{appfig:CN_noschem}
 \end{figure}
 
    \begin{figure}
 \centering
 \includegraphics[trim =0 0 0 0cm,width= 0.50\textwidth]{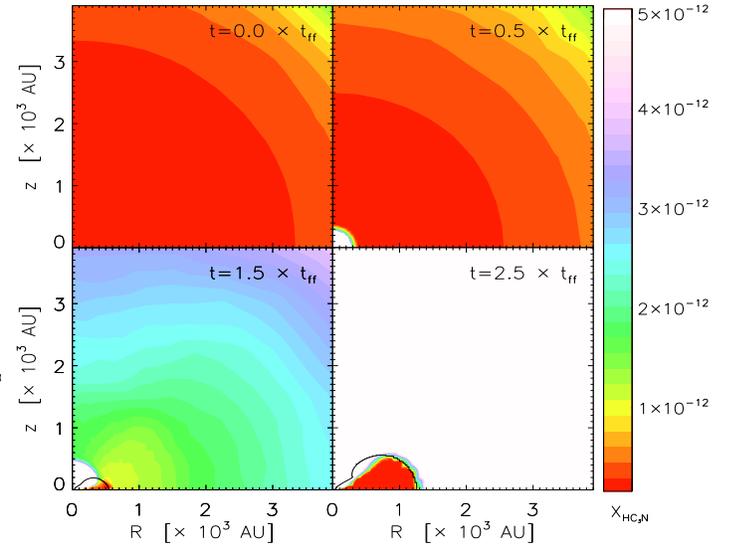}
 \caption{Same as for Fig. \ref{appfig:CO_noschem} but for \rm{HC$_{3}$N}}
 \label{appfig:HC3N_noschem}
 \end{figure}
 
     \begin{figure}
 \centering
 \includegraphics[trim =0 0 0 0cm,width= 0.50\textwidth]{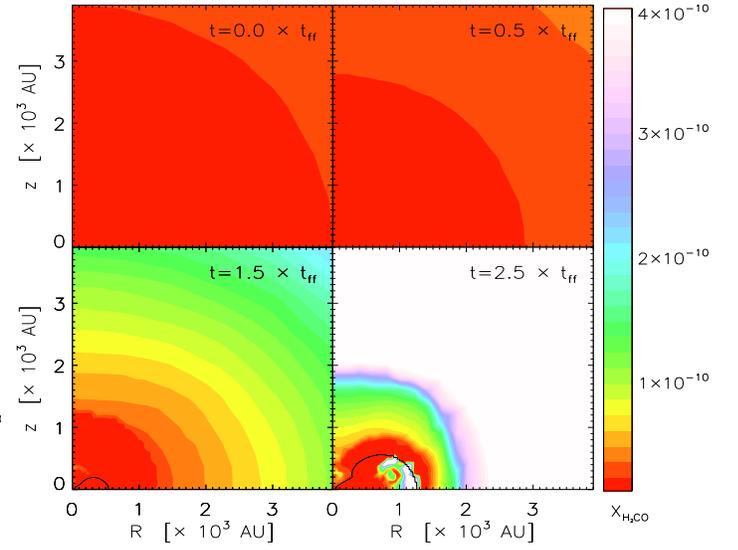}
 \caption{Same as for Fig. \ref{appfig:CO_noschem} but for \rm{H$_{2}$CO}}
 \label{appfig:H2CO_noschem}
 \end{figure}
 
     \begin{figure}
 \centering
 \includegraphics[trim =0 0 0 0cm,width=0.50\textwidth]{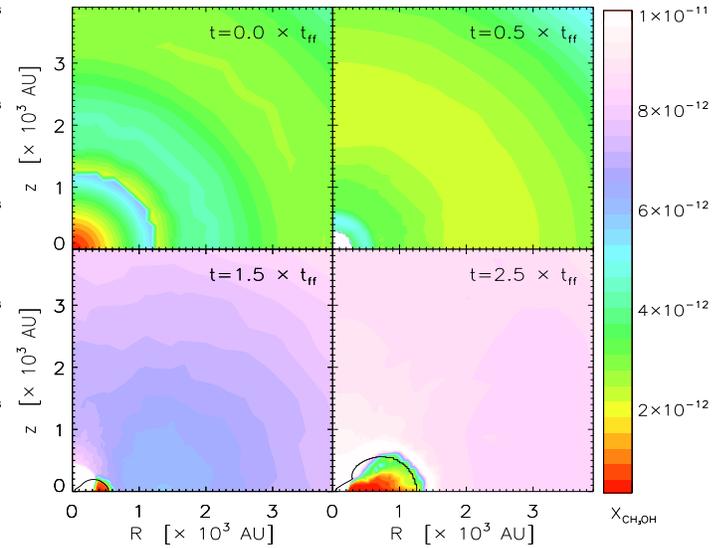}
 \caption{Same as for Fig. \ref{appfig:CO_noschem} but for \rm{CH$_{3}$OH}}
 \label{appfig:CH3OH_noschem}
 \end{figure}

 \end{appendix}

\end{document}